\documentclass[aps,preprint,showpacs,onecolumn]{revtex4}
\usepackage{amsmath,graphics}
\usepackage[next]{inputenc}
\usepackage[dvips]{epsfig}
\def\be{\begin{equation}}
\def\ee{\end{equation}}

\def\bi{\begin{itemize}}
\def\ei{\end{itemize}}
\def\bn{\begin{enumerate}}
\def\en{\end{enumerate}}
\def\bea{\begin{eqnarray}}
\def\eea{\end{eqnarray}}
\def\no{\nonumber}
\def\ba{\begin{array}}
\def\ea{\end{array}}
\def\bd{\begin{displaymath}}
\def\ed{\end{displaymath}}

 \newcommand{\Z}{\mathbf{Z}}
 \newcommand{\PP}{\mathbf{P}}

 \newcommand{\half}{\frac 1 2}

 \newcommand{\prima}{^\prime}

 \newcommand{\bbar}[1]{\bar{\bar {#1}}}

 \newcommand{\ket}[1]{|#1\rangle}
 \newcommand{\bra}[1]{\langle #1|}

 \newcommand{\bc}{{\bar{c}}}
 \newcommand{\bbc}{{\bbar{c}}}

 \newcommand{\rel}{,}

\begin{document}

\title[Topological Color Codes ...]{Topological Color Codes
and Two-Body Quantum Lattice Hamiltonians}

\author{M. Kargarian$^{1}$, H. Bombin$^{2}$ and M.A. Martin-Delgado$^{3}$}%

\address{$^{1}$Physics Department, Sharif University of Technology, Tehran 11155-9161, Iran \\
$^{2}$Department of Physics, Massachusetts Institute of Technology,
Cambridge, Massachusetts 02139, USA \\
$^{3}$Departamento de F\'{i}sica Te\'orica I, Universidad
Complutense, 28040 Madrid, Spain}

\begin{abstract}
Topological color codes are among the stabilizer codes with
remarkable properties from quantum information perspective. In this
paper we construct a lattice, the so called ruby lattice, with
coordination number four governed by a 2-body Hamiltonian. In a
particular regime of coupling constants, in a strong coupling limit,
degenerate perturbation theory implies that the low energy spectrum
of the model can be described by a many-body effective Hamiltonian,
which encodes the color code as its ground state subspace. Ground
state subspace corresponds to vortex-free sector. The gauge symmetry
$\mathbf{Z}_{2}\times\mathbf{Z}_{2}$ of color code could already be
realized by identifying three distinct plaquette operators on the
ruby lattice. All plaquette operators commute with each other and
with the Hamiltonian being integrals of motion. Plaquettes are
extended to closed strings or string-net structures.
Non-contractible closed strings winding the space commute with
Hamiltonian but not always with each other. This gives rise to exact
topological degeneracy of the model. Connection to 2-colexes can be
established via the coloring of the strings. We discuss it at the
non-perturbative level. The particular structure of the 2-body
Hamiltonian provides a fruitful interpretation in terms of mapping
to bosons coupled to effective spins. We show that high energy
excitations of the model have fermionic statistics. They form three
families of high energy excitations each of one color. Furthermore,
we show that they belong to a particular family of topological
charges. The emergence of invisible charges related to the
string-net structure of the model. The emerging fermions are coupled
to nontrivial gauge fields. We show that for particular 2-colexes,
the fermions can see the background fluxes in the ground state.
Also, we use Jordan-Wigner transformation in order to test the
integrability of the model via introducing of Majorana fermions. The
four-valent structure of the lattice prevents the fermionized
Hamiltonian to reduce to a quadratic form due to interacting gauge
fields. We also propose another construction for 2-body Hamiltonian
based on the connection between color codes and cluster states. The
corresponding 2-body Hamiltonian encodes cluster state defined on a
bipartite lattice as its low energy spectrum, and subsequent
selective measurements give rise to the color code model. We discuss
this latter approach along the construction based on the ruby
lattice.
\end{abstract}
\date{\today}


\maketitle
\tableofcontents
\section{Introduction}
\label{sect_I}

Topological color codes (TCC) are a whole class of models that
provide an instance of an interdisciplinary subject between Quantum
Information and the physics of Quantum Many-Body Systems.

Topological color codes were introduced \cite{topodistill} as a
class of topological quantum codes that allow a direct
implementation of the Clifford group of quantum gates suitable for
entanglement distillation, teleportation and fault-tolerant quantum
computation. They are defined on certain types of 2D spatial
lattices. They were extended to 3D lattices \cite{tetraUQC} in order
to achieve universal quantum computation with TCCs. This proposal of
topological quantum computation relies solely on the topological
properties of the ground state sector of certain lattice
Hamiltonians, without resorting to braiding of quasiparticle
excitations. In addition to these applications in Quantum
Information, topological color codes have also a natural application
in strongly correlated systems of condensed matter with topological
orders. In \cite{topo3D} was found that TCCs can be extended to
arbitrary dimensions, giving rise to topological orders in any
dimension, not just 2D. This is accomplished through the notion of
$D$-colexes, which are a class of lattices with certain properties
where quantum lattice Hamiltonians are defined. This corresponds to
a new class of exact models in D=3 and higher dimensions that
exhibit new mechanisms for topological order: i/ brane-net
condensation; ii/ existence of branyons; iii/ higher ground-state
degeneracy than other codes; iv/ different topological phases for
$D\geq 4$ etc. In what follows, we shall focus only on 2D lattices.

Physically, TCCs are exotic quantum states of matter with novel
properties. They are useful for implementing topological quantum
computation, but they have also an intrinsic interest by their own.
Then, a natural question arises as to how to implement
experimentally these new quantum systems by means of light, atoms or
some other means. This is a challenge since TCCs are formulated in
terms of Hamiltonians with many-body terms, the simplest having
6-body interactions in a hexagonal lattice. But the most common
interactions in nature are typically 2-body interactions.

There are several approaches to trying to solve this challenge,
depending on the type of scenario we envision to be in practice and
the practical rules we are supposed to be allowed to have at our
disposal.

Let us start first for what we may call a 'quantum control
scenario'. By this we simply mean that we are able to perform very
controllable quantum operations on our system that we have prepared
artificially. In particular, we suppose that we can perform quantum
measurements on the qubits and having ancilla qubits at will. Under
these circumstances, we can resort to cluster states \cite{br01}
and measurement-based quantum computation \cite{rb02, rbb03}. This
is because TCCs can be described by a certain cluster state
construction \cite{stat_colorcodes} within this scenario. Then, it
is possible to use a technique to obtain graph states as ground
states of 2-body qubit Hamiltonians \cite{Bartlett,nldb08}. We show
this construction in
 \ref{sect_appendix}.
However, this scenario is experimentally very demanding  and it is
left for the future when will it be achieved completely. Therefore,
it is convenient to seek other alternatives.

Thus, let us move onto a 'condensed matter scenario'. The
terminology is intended just to be illustrative, rather than
precise. In fact, the scenario goes beyond condensed matter and may
well be a quantum simulation of our system by means of engineering a
set of photons, atoms or the like. The important difference now is
that external measurements on the system, or ancilla qubits, are not
allowed in order to obtain the desired Hamiltonian for the TCCs. We
want to remain in a framework based on Hamiltonians with solely
2-body interactions \cite{AnyonicFermions}.

We have introduced  a new quantum 2-body Hamiltonian on a 2D lattice
with results that follow the twofold motivation concerning topics in
Quantum Information and Quantum Many-Body Systems:

\noindent i/ to achieve scalable quantum computation
\cite{NielsenChuang,rmp,landahl_etal09};

\noindent ii/ to perform quantum simulations with light, atoms and
similar available means
\cite{zoller05,jiang08,muller_etal08,duan03,spin_ladders_OL04,hartmann_etal06,
ol_review,suter_etal07,aktb08,bac09}.

This is so because, on one hand, the Hamiltonian system that we
introduce is  able to reproduce the quantum computational properties
of the topological color codes (TCC)
\cite{topodistill,topo3D,tetraUQC} at a non-pertubative level as
explained in Sect.\ref{sect_IV}. This is an important step towards
obtaining topological protection against decoherence in the quest
for scalability. On the other hand, the fact that the interactions
in the Hamiltonian appear as 2-body spin (or qubit) terms makes it
more suitable for its realization by means of a quantum simulation
based on available physical proposal with light and atoms.

In a framework of strongly correlated systems in Quantum Many-Body
Systems, one of the several reasons  for being interested in the
experimental implementation of this Hamiltonian system is because it
exhibits exotic quantum phases of matter known as topological
orders, some of its distinctive features being the existence of
anyons \cite{wilczek82,LM77,moore_read91}. In our everyday 3D world,
we only deal with fermions and bosons. Thus, exchanging twice a pair
of particles is a topologically trivial operation. In 2D this is no
longer true, and particles with other statistics are possible:
anyons. When the difference is just a phase, the anyons are called
abelian. Anyons are a signature of topological order (TO)
\cite{wen_book,wen_fqh}, and there are others as well:
\begin{itemize}
\item  there is an energy gap between the
ground state and the excitations;

\item topological degeneracy of the ground state subspace (GS);

\item this degeneracy cannot be lifted by local perturbations;

\item  localized quasiparticles as
excited states: anyons;

\item edge states;

\end{itemize}
\noindent etc.

These features re¯ect the topological nature of the system. In
addition, a signature of the TO is the dependence of that degeneracy
on topological invariants of the lattice where the system is de®ned,
like Betti numbers \cite{topo3D}.

But where do we find topological orders? These quantum phases of
matter are difficult to find. If we are lucky, we may find them on
existing physical systems such as the quantum Hall effect. But we
can also engineer suitable quantum Hamiltonian models, e.g., using
polar molecules on optical lattices
\cite{zoller05,jiang08,ol_review}, or by some other means. There are
methods for demonstrating topological order without resorting to
interferometric techniques \cite{interferometry_free}.

In this paper we present new results concerning the realization of
2-body Hamiltonians using cluster states techniques on one hand, and
without measurement-based computations on the other. In this latter
case, we present a detailed study of the set of integrals of motion
(IOM) in a 2-body Hamiltoinan, fermionic mappings of the original
spin Hamiltonian that give information about the physics of the
system and which complements previous results using bosonic mapping
techniques \cite{AnyonicFermions}.

This paper is organized as follows: in Sect. \ref{sect_II} we
present color codes as instances of topological stabilizer codes
with Hamiltonians based on many-body interacting terms and then
introduce the quantum Hamiltonian model based solely on 2-body
interactions between spin-$\frac{1}{2}$ particles. The lattice is
two-dimensional and has coordination number 4, instead of the usual
3 for the Kitaev model. It is shown in Fig.~\ref{ruby_lattice} and
it is called ruby lattice. In Sect. \ref{sect_III}, we describe the
structure of the set of exact integrals of motion (IOM) of the
2-Body model. We give a set of diagrammatic local rules that are the
building blocks to construct arbitrary IOMs. These include colored
strings and string-nets constant of motion, which is a distinctive
feature with respect to the Kitaev's model. In Sect. \ref{sect_IV},
we establish a connection between the original topological color
code and the new 2-Body color model. This is done firstly at a
non-perturbative level using the colored string integrals of motion
that are related with the corresponding strings in the TCC. Then,
using degenerate perturbation theory in the Green function
formalism, it is possible to describe a gapped phase of the 2-Body
color model that corresponds precisely to the topological color
code. In Sect. \ref{sect_V}, we introduce a mapping from the
original spin-$\frac{1}{2}$ degrees of freedom onto bosonic degrees
of freedom in the form of hard-core bosons which also carry a
pseudospin. This provides an alternative way to perform perturbation
theory and obtain the gapped phase corresponding to the TCC. It also
provides a nice description of low energy properties of the 2-Body
model and its quasiparticles. In Sect. \ref{sect_VI}, we introduce
another mapping based on spinless fermions which is helpful to
understand the structure of the 2-Body Hamiltonian and the presence
of interacting terms which are related to the existence of
stringnets constants of motion. Sect. \ref{sect_VII} is devoted to
conclusions and future prospects. \ref{sect_appendix} describes how
to obtain 2-Body Hamiltonians for topological color codes based on
cluster states and measurements using ancilla qubits.

\section{Quantum Lattice Hamiltonian with Two-Body Interactions}
\label{sect_II}

\subsection{Color Codes as Topological Stabilizers}

Some of the simplest quantum Hamiltonian models with topological
order can be obtained from a formalism based on the local stabilizer
codes borrowed from quantum error correction \cite{gottesman96} in
quantum information \cite{evans_stephens09a,evans_stephens09b}.
These are spin-$\half$ local models of the form \be H = -\sum_i S_i,
\ \ S_i\in \PP_n:=\langle i, \sigma^x_1, \sigma^z_1,\ldots,
\sigma^x_n, \sigma^z_n  \rangle. \label{stabilizer_Hamiltonian} \ee
where the stabilizer operators $S_i$ constitute an abelian subgroup
of the Pauli group $\PP_n$ of $n$ qubits, generated by the Pauli
matrices not containing $-1$. The ground state is a stabilizer code
since it satisfies the condition \be S_i\ket{{\rm GS}}=\ket{{\rm
GS}}, \forall i, \ee and the excited states of $H$ are gapped, and
correspond to error syndromes from the quantum information
perspective \be S_i\ket{\Psi}=-\ket{\Psi}.
\label{stabilizer_violation}\ee

\begin{figure}[h]
\begin{center}
\includegraphics[width=8 cm]{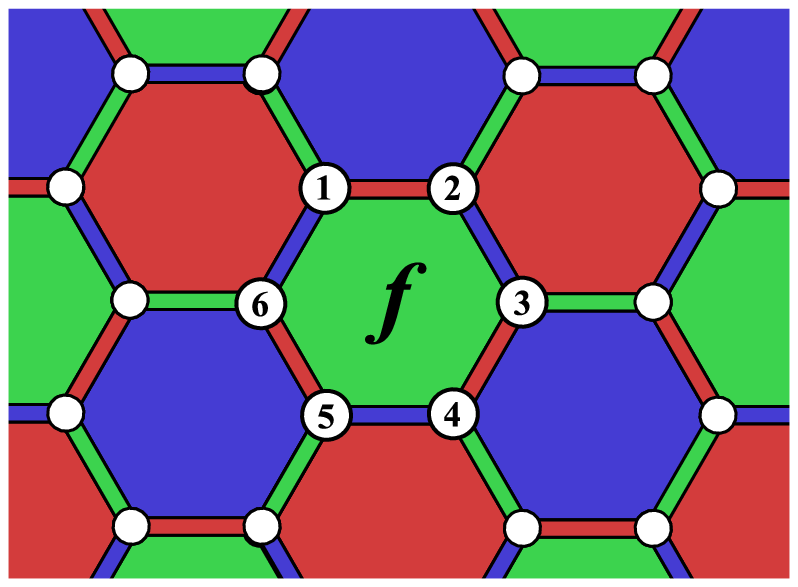}
\caption{\label{color_code} The hexagonal lattice is an example of
3-colorable lattice by faces, and also by edges. A topological color
code can be defined on it by associating two stabilizer operators
for each plaquette \eqref{color_stabilizers}.}
\end{center}
\end{figure}

\begin{figure}[h]
\begin{center}
\includegraphics[width=8 cm]{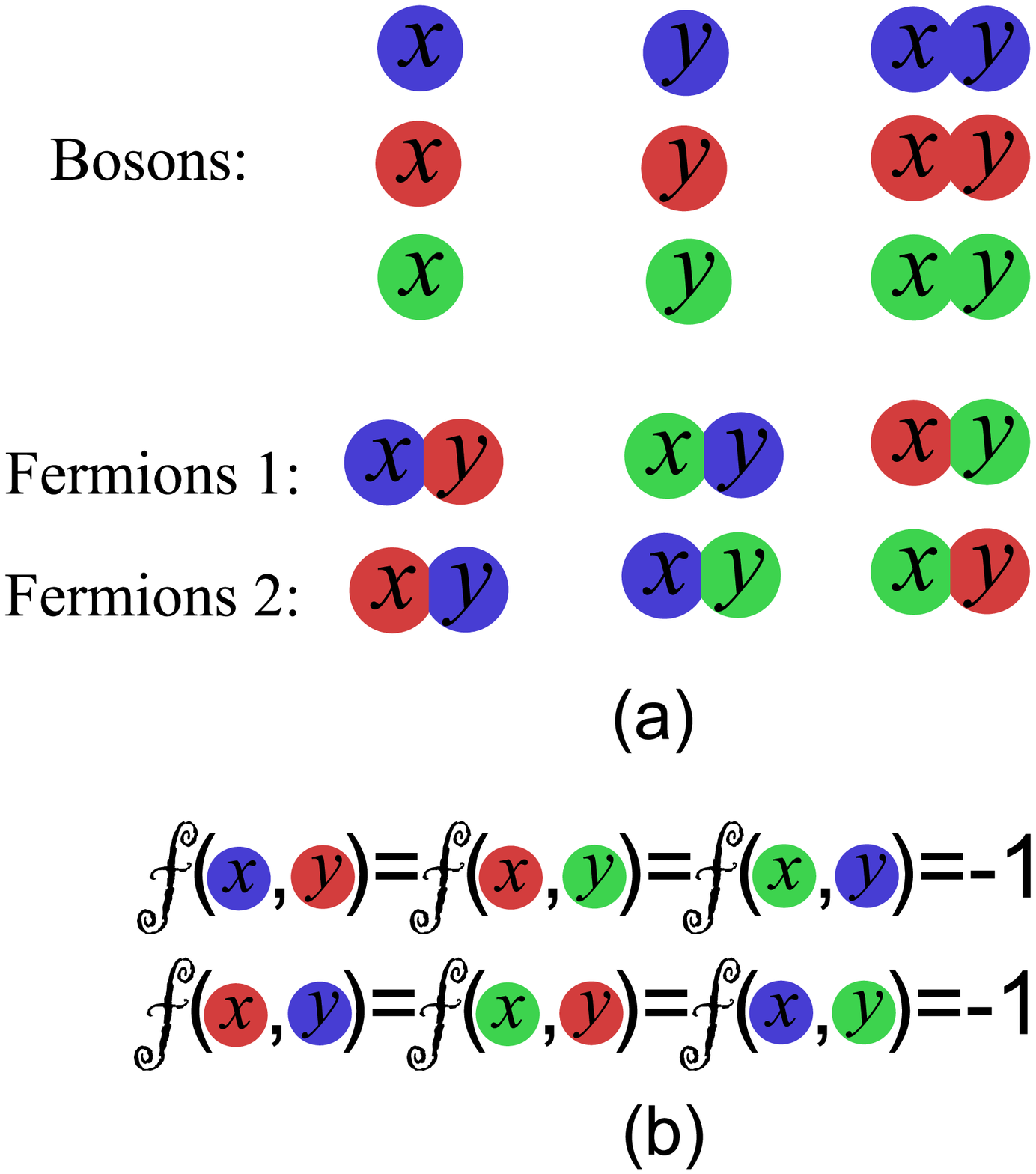}
\caption{(a) Classification of excitations for the topological color
code model \eqref{color_code_Hamiltonian}, nine bosons and two
families of fermions (b) The nontrivial phase arising from the
braiding of different charges. \label{topological_charges}}
\end{center}
\end{figure}

The seminal example of topological stabilizer codes is the toric
code \cite{kitaev97}.
There are basically two types of known topological stabilizer codes
\cite{landahl_etal09}. It is possible to study this type of
homological error correcting codes in many different situations and
perform comparative studies
\cite{homologicalerror,optimal,aps09,bullock_brennen07,surfaceVsColor,
planat_kibler,georgiev08,hu_etal08,georgiev06}. Topological color
code (TCC) is the another relevant example of topological stabilizer
codes, with enhanced computational capabilities
\cite{topodistill,tetraUQC,topo3D}. In particular, they allow the
transversal implementation of Clifford quantum operations. The
simplest lattice to construct them is a honeycomb lattice $\Lambda$
shown in Fig.\ref{color_code}, where we place a spin-$\half$ system
at each vertex. There are two stabilizer operators per plaquette:
\begin{equation}
\begin{split}
B_f^x&=\tau^x_1 \tau^x_2 \tau^x_3 \tau^x_4 \tau^x_5 \tau^x_6,\\
B_f^y&=\tau^y_1 \tau^y_2 \tau^y_3 \tau^y_4 \tau^y_5 \tau^y_6,
\end{split}
\label{color_stabilizers}
\end{equation}
\be H_{\rm cc} = -\sum_f (B^x_f + B^y_f),
\label{color_code_Hamiltonian} \ee
where $\tau^{\nu}$'s $(\nu=x,y)$ are usual Pauli operators. There
exist six kinds of basic excitations. To label them, we first label
the plaquettes with three colors: Notice that the lattice is
3-valent and has 3-colorable plaquettes. We call such lattices
2-colexes \cite{topo3D}. One can define color codes in any 2-colex
embedded in an arbitrary surface. There exists a total of 15
nontrivial topological charges as follows. The excitation at a
plaquette arises because of the violation of the stabilizer
condition as in \eqref{stabilizer_violation}. Consider a rotation
$\tau^{y}$ applied to a certain qubit. Since $\tau^{y}$ anticommutes
with plaquette operators $B^{x}_{f}$ of neighboring plaquettes, it
will put an excitation at corresponding plaquette. Similarly, if we
perform a $\tau^{x}$ rotation on a qubit, the plaquette operators
$B^{y}_{f}$ are violated. These are the basic excitations, two types
of excitations per each colored plaquette. Regarding the color and
type of basic excitations, different emerging excitations can be
combined. The whole spectrum of excitations is shown in
Fig.\ref{topological_charges}(a). Every single excitation is boson
by itself as well as the combination of two basic excitations with
the same color. They form nine bosons. However, excitations of
different color and type have semionic mutual statistics as in
Fig.\ref{topological_charges}(b). The excitations of different color
and type can also be combined. They form two families of fermions.
Each family of fermions is closed under fusion, and fermions from
different families have trivial mutual statistics. This latter
property is very promising and will be the source of invisible
charges as we will discuss in Sect.\ref{sect_V}. The anyonic charge
sectors are in one to one correspondence with the irreducible
representations (irreps) of the underlying gauge group, and the
fusion corresponds to decomposition of the tensor product of irreps.

We describe all above excitations in terms of representation of the
gauge group of the TCC. Before that, let us make a convention for
colors which will be useful for subsequent discussions. We refer to
colors by a bar operation $\bar{c}$ that transform colors cyclically
as $\mathrm{\bar{r}=g}$, $\mathrm{\bar{g}=b}$ and
$\mathrm{\bar{b}=r}$. The elements of the gauge group
$\mathbf{Z_{2}}\times \mathbf{Z_{2}}$ are $\{\mathrm{e,r,b,g}\}$.
Each excitation carries a topological charge. The corresponding
topological charge can be labeled by the pair $(q,\chi)$, where
$q\in \mathbf{Z_{2}}\times \mathbf{Z_{2}}$ and $\chi$ an irrep of
this group \cite{AnyonicFermions}. We label them as
$\chi_{e}(c)=\chi_{c}(c)=-\chi_{c}(\bar{c})=1$. Therefore, there are
nine bosons labeled by $(c,\chi_{e})$, $(e,\chi_{c})$ and
$(c,\chi_{c})$ and six fermions $(c,\chi_{\bar{c}})$ and
$(c,\chi_{\bar{\bar{c}}})$. Taking into account the vacuum with
trivial charge $(e,\chi_{e})$, color code has sixteen topological
charges or superselection sectors. Regarding the fusion process,
fusion of two charges $(q,\chi_{c})$ and $(q',\chi_{c'})$ give rises
to $(qq',\chi_{c}\chi_{c'})$ charge. Additionally, the braiding of
charge $(q,\chi_{c})$ around charge $(q',\chi_{c'})$ produces the
phase $\chi_{c}(q')\chi_{c'}(q)$. An excitation at a $c$-plaquette
has $(c,\chi_{e})$ charge if $-B^{x}=B^{y}=1$, $(e,\chi_{c})$ charge
if $B^{x} =-B^{y} =1$ and $(c,\chi_{c}$) charge if $B^{x}=B^{y}=-1$.

It is also possible to use both types of topological stabilizer
codes, either toric codes or color codes, to go beyond homological
operations. This corresponds to performing certain types of
operations called code deformations, which may alter the topology of
the surface allowing an extension of the computational capabilities
of these 2D codes
\cite{dennis_etal02,code_deformation07,rhg07,rh07,super09}.

Active error correction procedures are particularly interesting in
the case of topological stabilizer codes. They give rise to
connections with random statistical mechanical models like the
random bond Ising model for the toric code \cite{dennis_etal02} and
new random 3-body Ising models for color codes \cite{random_3Body}.
The whole phase diagram $p-T$ has been mapped out using Monte Carlo,
which in particular gives the value of the error threshold $p_c$.
This particular point can also be addressed using multicritical
methods\cite{ohthreshold09}. There is experimental realization of
topological error correction \cite{weibo_etal09}. Without external
active error correction, the effect of thermal noise is the most
challenging problem in toric codes
\cite{AlickiFH-Kitaev1,AlickiFH-Kitaev2,AlickiHHH-Kitaev,ipgap08b,ipgap08a,
bravyi_terhal}. Finite temperature effects of topological order in
color codes has also been studied \cite{finite_temperatureTCC}.

In all, the type of entanglement exhibited by topological color
codes is very remarkable
\cite{entanglementTCC,ng_etal09,pakman_parnachev08,
rico_briegel08,ecp09, dur_briegel07,vahid09}. A very illustrative
way to see this is using the connection of the ground state of
topological codes with standard statistical models by means of
projective measurements
\cite{ndb_07a,br_07,ndb_07b,stat_colorcodes,ndrb08,hndb08,cdnb08,cdbm08}.
For TCCs, this mapping yields  the partition function of a 3-body
classical Ising model on triangular lattices \cite{stat_colorcodes}.
This 3-body model is the same found in active error correcting
techniques \cite{random_3Body}, but without randomness since there
is no noise produced by external errors.
 This type of statistical mapping allows us to test that
different computational capabilities of color codes correspond to
qualitatively different universality classes of their associated
classical spin models. Furthermore,  generalizing these statistical
mechanical models for arbitrary inhomogeneous and complex couplings,
it is possible to study a measurement-based quantum computation with
a color code state and we find that their classical simulatability
remains an open problem. This is in sharp contrast with toric codes
which are classically simulable within this type of scheme
\cite{br_07}.

\subsection{The Model}

In nature, we find that interactions are usually 2-body
interactions. This is because interactions between particles are
mediated by exchange bosons that carry the interactions
(electromagnetic, phononic, etc.) between two particles.

The problem that arises is that for topological models, like the
toric codes and color codes, their Hamiltonians have many-body terms
\eqref{color_code_Hamiltonian}. This could only achieved by finding
some exotic quantum phase of nature, like FQHE, or by artificially
enginering them somehow.

Here, we shall follow another route: try to find a 2-body
Hamiltonian on a certain 2D lattice such that it exhibits the type
of topological order found in toric codes and color codes. In this
way, their physical implementation looks more accessible.

In fact, Kitaev \cite{honeycomb} introduced a 2-body model in the
honeycomb lattice that gives rise to an effective toric code model
in one of its phases. It is a 2-body spin-$\half$ model in a
honeycomb lattice with one spin per vertex, and simulations based on
optical lattices have been proposed \cite{duan03}.

The model features plaquette and strings constants of motion.
Furthermore, it is exactly solvable, a property that is related to
the  3-valency of the lattice where it is defined
\cite{honeycomb,Feng_et_al07,Yao_Kivelson07,baskaran_etal07,Yang_et_al07,
si_yu08,mandal_surendran09}. It shows emerging free fermions in the
honeycomb lattice. If a magnetic field is added, it contains a
non-abelian topological phase (although not enough for universal
quantum computation). Interestingly enough, another regime of the
model gives rise to a 4-body model, which is precisely an effective
toric code model. A natural question arises: Can we get something
similar for color codes? We give a positive answer in what follows.

Motivated by these physical considerations related to a typical
scenario in quantum many-body physics, either condensed matter, AMO
physics or the like, we will seek a quantum spin Hamiltonian with
the following properties:

\noindent i/ One of its phases must be the TCC.

\noindent ii/ To have two sets of plaquette operators generating a
$\Z_2\times\Z_2$ local, i.e. gauge, symmetry.

\noindent iii/ To have string-nets and colored strings IOM as in the
TCC, but in all coupling regimes.

Thus, the reasons behind demanding these properties are to guarantee
that the sought after model will host the TCC. For instance,
property i/ means that we must be able to generate the 2D color code
Hamiltonian consistently at some lowest order in perturbation theory
(PT). This we shall see in Sect.\ref{Green_Function}. Likewise,
properties ii/ and iii/ are demanded in order to have the
fundamental signatures regarding gauge symmetry and constants of
motions associated with TCCs. Notice that we have not demanded that
the model be exactly solvable. This is a mathematical requisite,
rather than physical. We leave the door open for considering larger
classes of models beyond exactly solvable models, which may be very
interesting and contain new physics. For example, according to those
properties, it would be possible to have models with a number of
IOMs that scales linearly with $N$, the number of spins or qubits.
Thus, the Kitaev model has a number of IOMs of $\half N$.

Our purpose is to present first the 2-body quantum Hamiltonian in 2D
\cite{AnyonicFermions}, and then to analyze diverse possible
mappings in later sections, like using bosonic and fermionic degrees
of freedom. The analysis of the set of IOMs will play also a crucial
role in the understanding of our model as we shall see in
Sect.\ref{sect_III}

\begin{figure}[h]
\begin{center}
\includegraphics[width=10 cm]{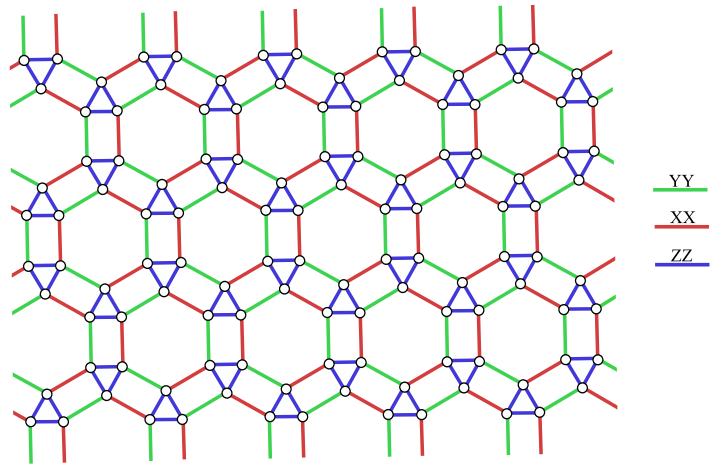}
\caption{\label{ruby_lattice} A lattice with coordination number 4
where the 2-body quantum lattice Hamiltonian for the color codes is
defined according to spin-spin interactions coded by the colors of
the links, as in \eqref{2-bodyTCC}. A plaquette can be distinguished
by an inner hexagon, an outer hexagon and six blue triangles between
them.}
\end{center}
\end{figure}

It is a 2-body spin-1/2 model in a 'ruby' lattice as shown in
Fig.\ref{ruby_lattice}. We place one spin per vertex. Links come in
3 colors, each color representing a different interaction.

\be H = -\sum_{\langle i,j\rangle} J_w \sigma^w_i \sigma^w_j, \ \ \
\ \ w = \begin{cases}
x, & \text{ red links} \\
y, & \text{ green links} \\
z, & \text{ blue links}
\end{cases}
\label{2-bodyTCC} \ee

For a suitable coupling regime, this model gives rise to an
effective color code model. Furthermore, it exhibits new features,
many of them not present in honeycomb-like models:

\begin{itemize}

\item Exact topological degeneracy in all coupling regimes
($4^g$ for genus $g$ surfaces).

\item String-net integrals of motion.

\item Emergence of 3 families of strongly interacting fermions
with semionic mutual statistics.

\item $\Z_2\times\Z_2$ gauge symmetry.
Each family of fermions sees a different $\Z_2$ gauge subgroup.

\end{itemize}

\section{String Operators and Integrals of Motion}
 \label{sect_III}

\begin{figure}[h]
\begin{center}
\includegraphics[width=10 cm]{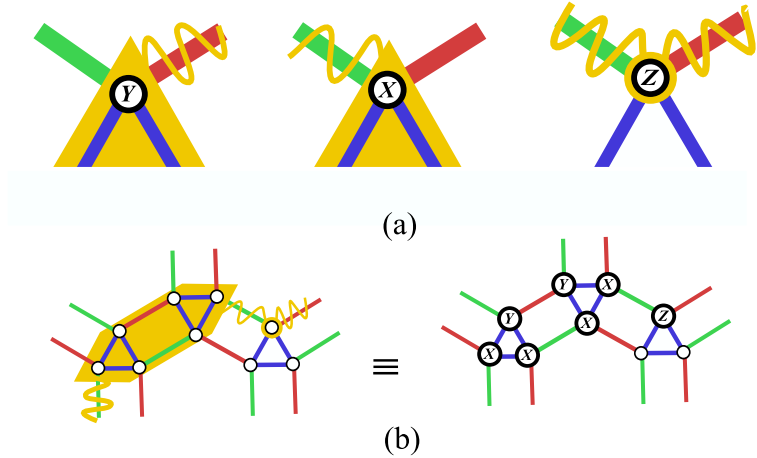}
\caption{\label{plaquette_IOM} (a) A diagrammatic representation of
the local structure of the integrals of motion of the 2-body
Hamiltonian \eqref{2-bodyTCC}. The colored links represent different
spin-spin interactions. (b) An example of contribution of Pauli
operators in a string.}
\end{center}
\end{figure}

We can construct integrals of motion (IOM), $I\in\PP_n, [H_{\rm
cc},I] = 0$, following a pattern of rules assigned to the vertices
of the lattice, as shown in Fig.\ref{plaquette_IOM}. These rules are
constructed to attach a Pauli operator of type $\sigma^x_i$,
$\sigma^y_i$ or $\sigma^z_i$ to each of the vertices $i$. The lines
around the vertices, either wavy lines or direct lines, are pictured
in order to join them along paths of vertices in the lattice that
will ultimately translate into products of Pauli operators, which
will become IOMs. Clearly, $\sigma^z_i$ operators are distinguished
from the rest. The contribution of each qubit in the string operator
is determined in terms of how it appears in the string. Its
contribution may be determined by the outgoing red and green links
which have the qubit as their end point in the string. In this case
the $\sigma^{x}$ or $\sigma^{y}$ Pauli operators contribute in the
string IOM . If a typical qubit crossed only by a wavy line as shown
in Fig.\ref{plaquette_IOM}(a), it contributes a $\sigma^{z}$ Pauli
operator in the string. To have a clear picture of string operators,
a typical example has also been shown in Fig.\ref{plaquette_IOM}(b).
Part of string is shown on the left and its expression will be the
product of Pauli operators which have been inserted in open circles
on the right. With such definitions for string operators and their
supports on the lattice, now we turn on to analyze the relevance of
strings to the model. In particular, we will construct elementary
string operators with the local symmetry of the model. Therefore, in
this way we are representing the local structure of the IOMs of our
2-body Hamiltonian \eqref{color_code_Hamiltonian}. We will
illustrate them with several examples of increasing complexity.
The ground state of a lattice model described by the Hamiltonian
\eqref{color_code_Hamiltonian} is a superposition of all closed
colored strings. Indeed, it is invariant under any deformations of
colored strings as well as splitting of a colored string into other
colors. In other words, the ground state is a string-net condensed
state and supports topological order. The gauge group related to
this topological order is $\mathbf{Z_{2}} \times \mathbf{Z_{2}}$.
Such symmetry of topological color code can be realized via defining
a set of closed string operators on the ruby lattice. We shall
verify the gauge symmetry by identifying a set of string operators
on the lattice of Fig.\ref{ruby_lattice}.

\begin{figure}[h]
\begin{center}
\includegraphics[width=11 cm]{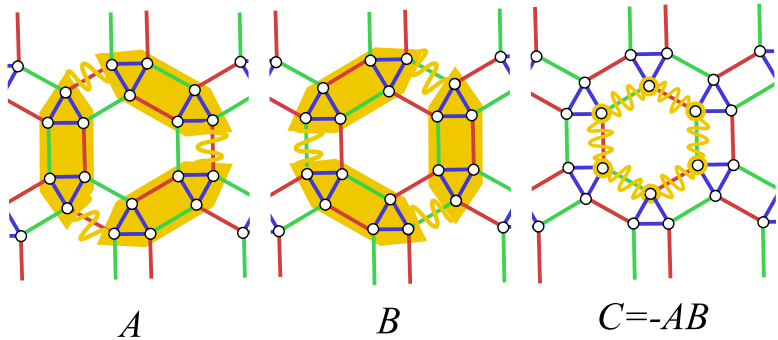}
\caption{\label{plaquette_IOM_Independent} Schematic drawing of the
plaquette IOMs according to the local rules in
Fig.\ref{plaquette_IOM}. There are 3 IOMs denoted as $A,B,C$, but
only 2 of them are independent. This corresponds to the symmetry
$\Z_2\times\Z_2$ of the model.}
\end{center}
\end{figure}

Let us start by constructing the elementary string IOM as shown in
Fig.\ref{plaquette_IOM_Independent}. They are denoted as $I=A,B,C$.
They are closed since they have not endpoints left. The elementary
closed strings are plaquettes. By a plaquette we mean an inner
hexagon and an outer hexagon with six triangles in between. For a
given plaquette it is possible to attach three string operators. For
each closed string, the contribution of Pauli operators are
determined based on outgoing red and green links or wavy lines as in
Fig.\ref{plaquette_IOM}. Let $V_{f}$ stand for a set of qubits on a
plaquette. Note that each plaquette contains $18$ qubits
corresponding to six triangles around it. For first plaquette
operator in Fig.\ref{plaquette_IOM_Independent} we can write its
explicit expression in terms of Pauli matrices as
$S^{A}_{f}=\prod_{i\in V_{f}}\sigma^{\nu}_{i}$, where $f$ denotes
the plaquette and $\nu=x, y$, depending on outgoing red or green
links, respectively. Similarly the second plaquette operator has an
expression as $S_{f}^{B}=\prod_{i\in V_{f}}\sigma^{\nu}_{i}$. The
third string is only a closed wavy string which coincides to the
inner hexagon of the plaquette. It's expression is
$S_{f}^{C}=\prod_{i\in V_{h}}\sigma^{z}_{i}$, where $V_{h}$ stands
for six qubits on the inner hexagon. The three closed strings
described above are not independent. Using the Pauli algebra, it is
immediate to check that they satisfy
$S^{C}_{f}=-S_{f}^{A}S_{f}^{B}$. Thus, there exist 2 independent
IOMs per plaquette: this is the $\Z_2\times\Z_2$ local symmetry of
the model Hamiltonian \eqref{2-bodyTCC}.

Plaquette operators commute with each other and with any other IOM.
If a IOM corresponds to a nontrivial cycle $c$, it is possible to
find another IOM that anticommutes with it, namely one that
corresponds to a cycle that crosses once $c\prima$ . Thus, IOMs
obtained from nontrivial cycles are not products of plaquette
operators.

Each string operator squares identity since we are working with
qubits. Plaquette operators corresponding to different plaquettes
commute with each other and also with terms in Hamiltonian in
\eqref{2-bodyTCC} since they share in zero or even number of qubits.
Therefore, the closed strings with the underlying symmetry obtained
above define a set of integrals of motion. The number of integrals
of motions is exponentially increasing. Let $3N$ be the total number
of qubits, so the number of plaquettes will be $\frac{N}{2}$.
Regarding to the gauge symmetry of the model, the number of
independent plaquette operators is $N$. This implies that there are
$2^{N}$ integrals of motion and allow us to divide the Hilbert space
into $2^{N}$ sectors being eigenspaces of plaquette operators.
However, for closed manifold, for example a torus, all plaquette
operators can not be independently set to $+1$ or $-1$ because they
are subject to some constraint. All other closed string operators
that are homologous to zero, i.e. they are homotopic to the boundary
of a plaquette, are just the product of these elementary plaquette
operators. It is natural that all of them are topologically
equivalent up to a deformation and commute with the Hamiltonian of
the model.
\begin{figure}[h]
\begin{center}
\includegraphics[width=11 cm]{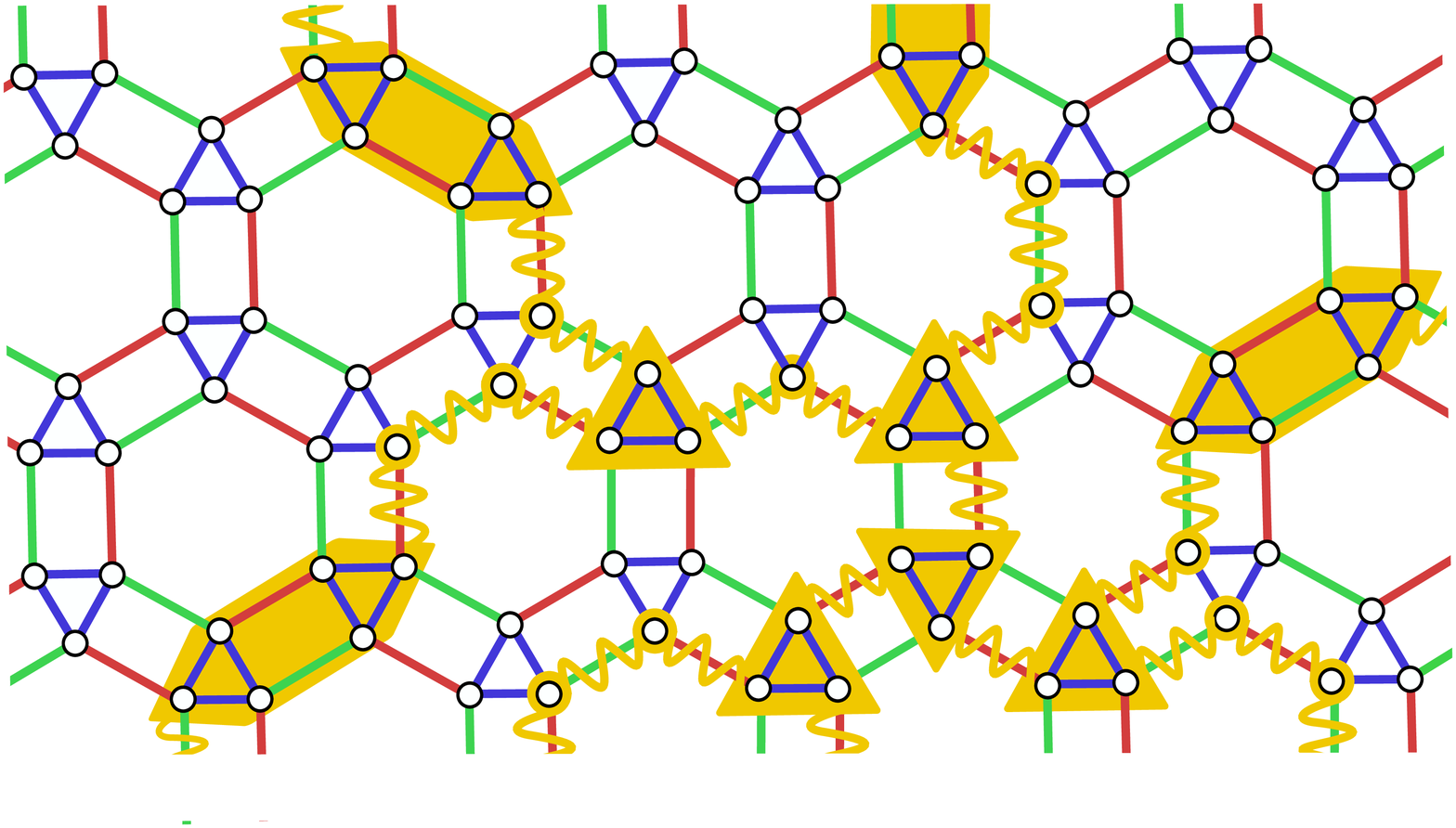}
\caption{\label{stringnet_IOM} An example of a stringnet IOM. Notice
the presence of branching points located around blue triangles of
the lattice. This is a remarkable difference with respect to
honeycomb models like the Kitaev model.}
\end{center}
\end{figure}

The most general configuration that we may have is shown in
Fig.\ref{stringnet_IOM}. We call them string-nets IOM since in the
context of our model, they can be thought of as the string-nets
introduced to characterize topological orders \cite{levinwen05}. The
key feature of these IOMs is the presence of branching points
located at the blue triangles of the lattice. This is remarkable and
it is absent in honeycomb 2-body models like the Kitaev model. When
the string-nets IOM are defined on a simply connected piece of
lattice they are products of plaquette operators. More generally,
they can be topologically non-trivial and independent of plaquette
operators.

\begin{figure}[h]
\begin{center}
\includegraphics[width=12 cm]{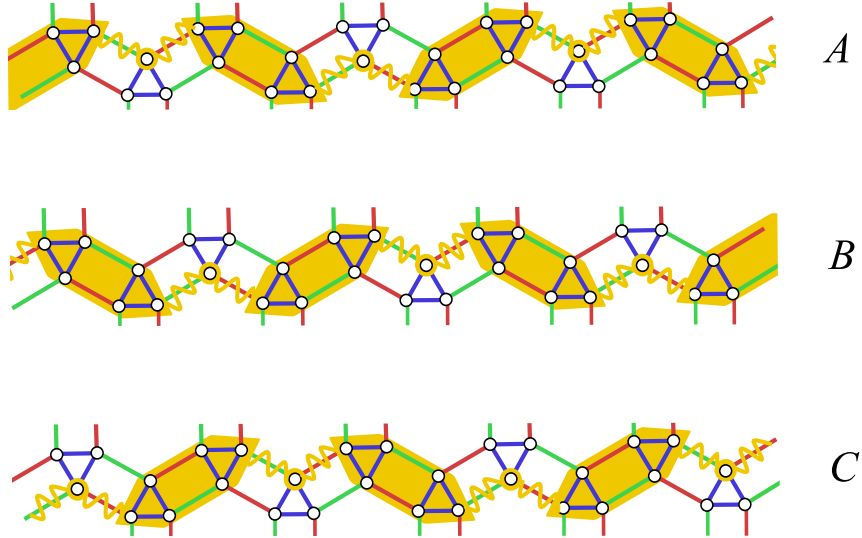}
\caption{\label{strings_IOM} Examples of standard string
configurations of IOMs, i.e., without branching points. For each
path, we can in principle make 3 different assignments of IOMs, but
again only 2 of them are independent as with plaquette IOMs. This is
another manifestation of the $\Z_2\times\Z_2$ symmetry  of the
model.}
\end{center}
\end{figure}

As a special case of IOMs we have string configurations, i.e., paths
without branching points. They correspond to the different homology
classes of the manifold where the lattice is embedded , and are
needed for characterization of the ground state manifold. Some
examples are shown in Fig.\ref{strings_IOM}. They may be open or
closed, depending on whether they have endpoints or not,
respectively. Strings IOM are easier to analyze. String-nets IOM are
products of strings IOM. For a given path, there exist 3 different
strings IOM. These are denoted as $A,B,C$ in Fig.\ref{strings_IOM}.
We must introduce generators for the homology classes defining
closed manifold. Homology classes of the torus are determined by
realizing two nontrivial loops winding around the torus. In the
Kitaev's model there are only two independent such nontrivial closed
loops. However, the specific construction of the lattice and
contribution of the color make it possible to define for each
homology class of the torus two independent nontrivial loops. These
closed strings are no longer combination of plaquettes defined
above. Let $S^{A}_{\mu}$ stand for such string, where $A$ and $\mu$
denote the type and homology class of the string. For each homology
class of the manifold we can realize three different types of string
operators depending on how the vertices of the lattice are crossed
by the underlying string. Each qubit crossed by the string
contributes a Pauli operator according to the rules in
Fig.\ref{plaquette_IOM}. Again, using Pauli algebra we can see that
only two of them are independent, as with the plaquette IOMs.  \bea
\label{symmetry_string}
(-1)^{\frac{t}{2}}S^{A}_{\mu}S^{B}_{\mu}S^{C}_{\mu} = 1,\eea where
$t$ is the number of triangles on the string. To distinguish
properly the three types we have to color the lattice.  We could
already use the colors to label strings. Strings are then red, green
or blue. This is closely related to the topological color code
\cite{topodistill,AnyonicFermions}. The latter relation shows that
each string can be constructed of two other homologous ones, which
is exactly the expression of the $\mathbf{Z_{2}} \times
\mathbf{Z_{2}}$ gauge symmetry. Each non-contractible closed string
operator of any homology commutes with all plaquette operators and
with terms appearing in the Hamiltonian, so they are constant of
motions. But, they don't always commute with each other. In fact, if
the strings cross once then  \bea \label{commue_string} \left [
S^{r}_{\mu} , S^{r}_{\nu} \right]=0 ,\eea but  \bea
\label{anticommue_string} \left \{ S^{r}_{\mu} , S^{r'}_{\nu}
\right\}=0.\eea This latter anticommutation relation is a source of
exact topological degeneracy \cite{Oshikawa_degeneracy} of the model
independent of phase we are analyzing it.

\section{A Gapped Phase: The Topological Color Code}
\label{sect_IV}

\subsection{Non-Perturbative Picture}

\begin{figure}
\begin{center}
\includegraphics[width=11cm]{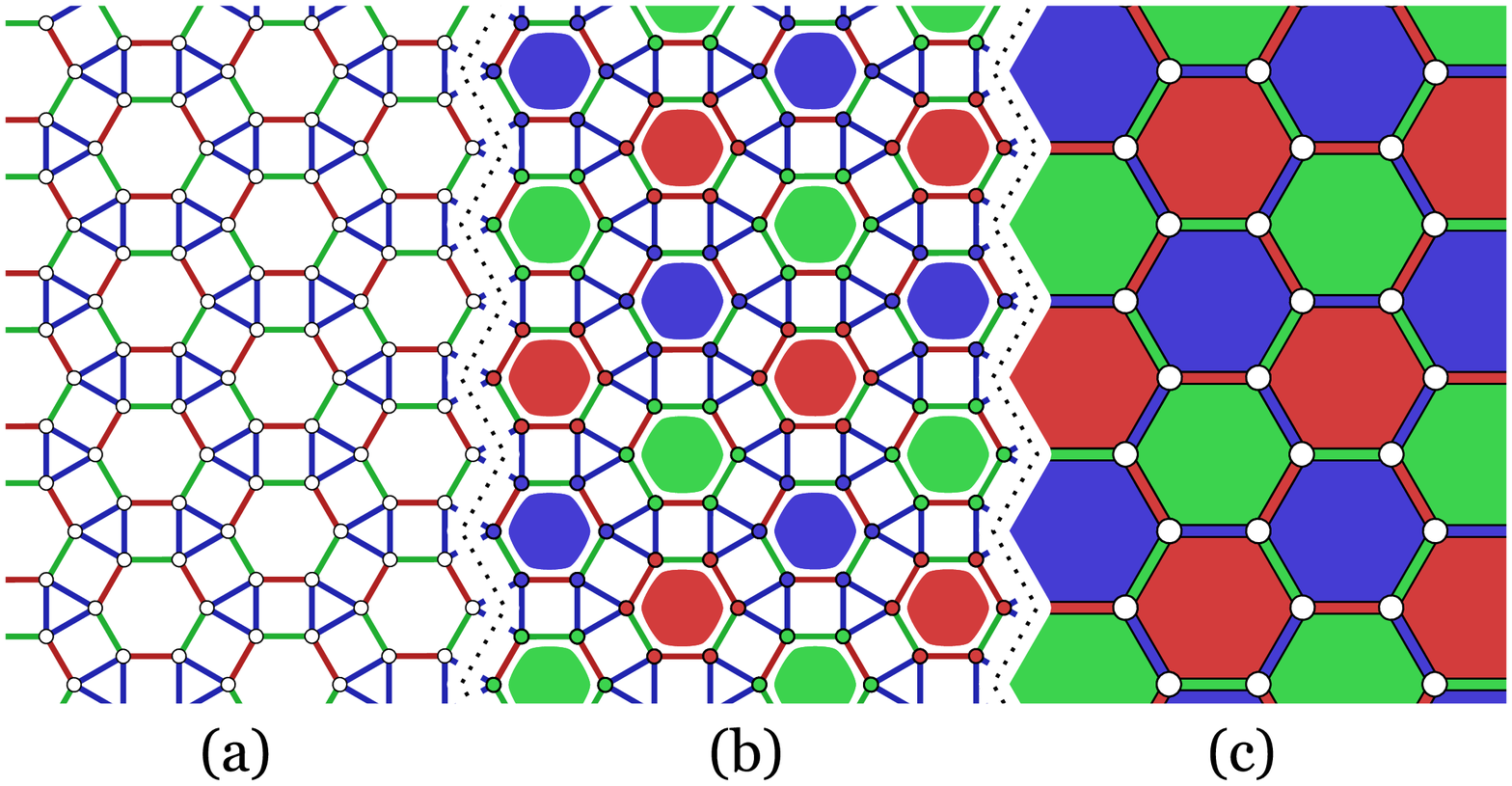} \caption{(color online)
The three stages showing the emergence of the topological color
code: (left) the original lattice for the 2-body Hamiltonian
\eqref{2-bodyTCC}. The colors in the links denote the type of
spin-spin interactions; (middle) a different coloring of the lattice
is introduced based on the property that the hexagons are
3-colorable, as well as the vertices; (right) the hexagonal lattice
obtained by shrinking to a point the blue triangles of the original
lattice, which become sites in the final hexagonal lattice. This
corresponds to the strong coupling limit in \eqref{triangle_limit}.
} \label{lattice_letter}
\end{center}
\end{figure}

In this subsection we discuss the ruby lattice is connected to the
2-colex even at the non-perturbative level. Then, in the subsequent
sections we verify it using quantitative methods. From the previous
discussion on IOMs, we have already seen a connection with the
topological color codes. Now, we want to see how different strings
introduced above are related to coloring of the lattice. To this
end, consider the closed strings $A, B, C$ in
Fig.\ref{plaquette_IOM_Independent}. The closed strings $A$ and $B$
can be visualized as a set of \emph{red} and \emph{green} links,
respectively. With such visualization, we put forward the next step
to color the inner hexagons of the ruby lattice: a colored link, say
red, connect the red inner hexagons. Accordingly, other inner
hexagons and links can be colored, and eventually we are left with a
colored lattice. The emergence of the topological color code is
beautifully pictured in Fig.\ref{lattice_letter}. Geometrically, it
corresponds to shrinking the blue triangles of the original lattice
into points, which will be referred as sites of a new emerging
lattice, see Fig.\ref{lattice_letter} (left). Thus, we realize that
the inner hexagons and vertices of the model are colorable, see
Fig.\ref{lattice_letter} (middle): if we regard blue triangles as
the sites of a new lattice, we get a honeycomb lattice, see
Fig.\ref{lattice_letter} (right). In fact, the model could be
defined for any other 2-colex, not necessarily a hexagonal lattice.

\begin{figure}
\begin{center}
\includegraphics[width=12cm]{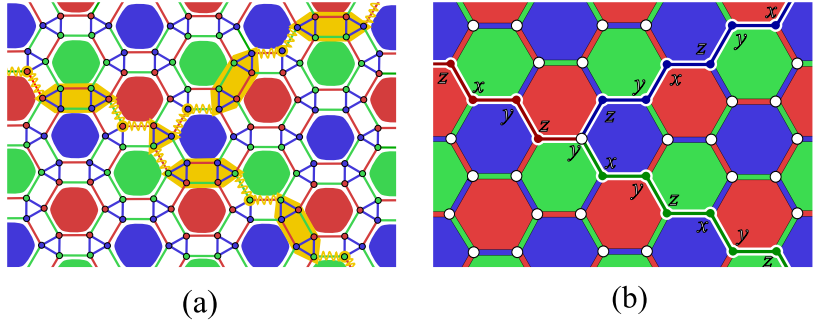} \caption{(color online) An illustraion
of correspondence between (a) strings on the ruby lattice,
corresponding to a non-perturbative picture, and (b)colored strings
on the effective honeycomb lattice.} \label{stringnet_effective}
\end{center}
\end{figure}

Connection to the 2-colexes can be further explored by seeing how
strings on the ruby lattice correspond to the colored strings on the
effective honeycomb lattice. To this end, consider a typical
string-net on the ruby lattice as shown in
Fig.\ref{stringnet_effective}(a). This corresponds to a
non-perturbative picture of the IOMs of the model. The fat parts of
the string-net connect two inner hexagons with the same color. In
this way, the corresponding string-net on the effective lattice can
be colored as in Fig.\ref{stringnet_effective}(b). The color of each
part of the string-net on the effective honeycomb lattice is
determined by seeing which colored inner hexagons on the ruby
lattice it connects. Three colored strings cross each other at a
branching point, and its expression in terms of Pauli matrices of
sites are given by product of Pauli operators written adjacent to
the sites. How they are determined, will be clear soon.

It is possible to use colors to label the closed strings on the
honeycomb lattice. Before that, let us use a notation for Pauli
operators acting on effective spins of honeycomb lattice
$\tau^{\alpha}$, where $\alpha=x,y,z$. We indicate the labels
$\alpha$ as $c|c:=z, \bar{c}|c:=x, \bar{\bar{c}}|c:=y$, where we are
using a bar operator. To each \emph{c}-plaquette, we attach three
operators each of one color. Let $B^{c'}_{f}$ denotes such
operators, where low and up indices stand for \emph{c}-plaquette $f$
and color of the closed string attached to the plaquette,
respectively. With these notations, the plaquette operators read as
follows:\bea B^{c'}_{f}=\prod_{v\in f}\tau_{v}^{c'|c},\eea where the
product runs over all vertices of the \emph{c}-plaquette $f$ in the
honeycomb lattice in Fig.\ref{lattice_letter}(c). Thus, we can write
the explicit expression of operators as follows:\bea \nonumber &&
B^{x}_{f}=B^{\bar{c}}_{f}=\prod_{v\in f}\tau_{v}^{\bar{c}|c}\\
\nonumber &&
B^{y}_{f}=B^{\bar{\bar{c}}}_{f}=\prod_{v\in f}\tau_{v}^{\bar{\bar{c}}|c}\\
\label{plaquette_honycomb1}&& B^{z}_{f}=-B^{c}_{f}=-\prod_{v\in
f}\tau_{v}^{c|c}.\eea All these plaquette operators are constant of
motions. Again, We can realize the gauge symmetry $\mathbf{Z_{2}}
\times \mathbf{Z_{2}}$ through the relation
$B^{x}_{f}B^{y}_{f}B^{z}_{f}=1$. On a compact manifold, for example
on the torus, all plaquettes are not independent. They are subject
to the following constraint: \bea \label{constraint} \prod_{f\in
\Lambda} B^{c}_{f}=(-1)^{N/2},\eea where the product runs over all
plaquettes $f$ in the lattice $\Lambda$, and $N$ is the total number
of plaquettes.

\begin{figure}
\begin{center}
\includegraphics[width=9cm]{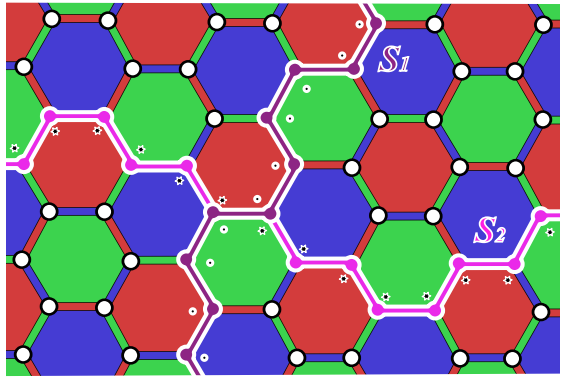} \caption{(color online) A piece of
effective lattice. The strings $S_{1}$ and $S_{2}$ correspond to
different homology classes of the manifold. Their expression in
terms of Pauli operators are given by their associated color and the
fact that how they turn on plaquettes on the lattice.}
\label{string_honeylattice}
\end{center}
\end{figure}

We can also realize noncontractible strings on the effective lattice
which are rooted in the topological degeneracy of the model. They
are just the IOMs in Fig.\ref{strings_IOM} when reduced on the
effective honeycomb lattice. Once the inner hexagons of ruby lattice
are colored, they correspond to colored strings as in
Fig.\ref{stringnet_effective}. Let $S^{c}_{\mu}$ stands for such
string, where indices $\mu$ and $c$ denote the homology and color of
the string, respectively. This string operator is tensor product of
Pauli operators of qubits lying on the string. Namely, the string
operator is \bea \label{IOM_honey} S^{c}_{\mu}=\prod_{v}
\tau^{c'|c}_{v}.\eea The contribution of each qubit is determined by
the color of the hexagon that the string turns on it, see
Fig.\ref{string_honeylattice}. For example in the string $S_{1}$
shown in Fig.\ref{string_honeylattice}, the color of the plaquettes
appearing in \eqref{IOM_honey} marked by light circles. With this
definition for string operators, the contribution of Pauli operators
in the string-net on the effective lattice in
Fig.\ref{stringnet_effective}(b) are reasonable. Non-contractible
colored strings are closely related to the topological degeneracy of
the model, since they commute with color code
Hamiltonian\eqref{color_code_Hamiltonian} being integrals of motion,
but not always with each other. In fact, two strings differing in
both homology and color anticommute, otherwise they commute. For
example let us consider two non-contractible closed strings $S_{1}$
and $S_{2}$ corresponding to different homologies of the torus. As
shown in Fig.\ref{string_honeylattice}, they share two qubits.
First, suppose both strings are of blue type. The contribution of
Pauli operators of these two qubits in string $S_{1}$ is
$\tau^{y}_{1}\tau^{x}_{2}$, while for string $S_{2}$ the
contribution is $\tau^{x}_{1}\tau^{y}_{2}$ implying
$[S^{b}_{1},S^{b}_{2}]=0$. Then, let $S_{2}$ be a green string. In
this case the contribution of qubits will be
$\tau^{y}_{1}\tau^{z}_{2}$, which explicitly shows that
$\{S^{b}_{1},S^{g}_{2}\}=0$. The interplay in
\eqref{symmetry_string} can be translated into an interplay between
color and homology as follows. \be
(-1)^{\frac{s}{2}}S^{c}_{\mu}S^{\bar{c}}_{\mu}S^{\bar{\bar{c}}}_{\mu}
= 1,\ee where $s$ is the number of sites on the string. This
interplay makes the ground state subspace of the color code model be
a string-net condensed phase.
\subsection{Degenerate Perturbation Theory: Green Function Formalism}
\label{Green_Function}

In this subsection we put the above correspondence between original
2-body lattice Hamiltonian and color code model on a quantitative
level.  In fact, there is a regime of coupling constants in which
one of the phases of the 2-body Hamiltonian reproduces the TCC
many-body structure and physics. In particular, we show that this
corresponds to the following set of couplings in the original 2-body
Hamiltonian \be J_{x},J_{y},J_{z}>0, ~~~J_{x},J_{y}\ll
J_{z},\label{triangle_limit}\ee that is, a strong coupling limit in
$J_{z}$. The topological color code effectively emerges in this
coupling regime. This can be seen using degenerate perturbation
theory in the Green function formalism. Let $H=H_{0}+V$ be a
Hamiltonian describing a physical system with two-body interaction,
and we regard the $\|V\|$, the norm of $V$, be very small in
comparison with the spectral gap of \emph{unperturbed} $H_{0}$. We
also suppose that $H_{0}$ has a degenerate ground-subspace which is
separated from the excited states by a gap $\Delta$. The effect of
$V$ will be to break the ground state degeneracy in some order of
perturbation. Now the interesting question is whether it is possible
to construct an effective Hamiltonian, $H_{\mathrm{eff}}$,  which
describes the low energy properties of the perturbed Hamiltonian
$H$. The effective Hamiltonian arises at orders of perturbation that
break the ground state degeneracy. From the quantum information
perspective the Hamiltonian $H$ acts on the physical qubits while
the effective Hamiltonian acts on the logical qubits projected down
from the physical qubits.

We will clarify that how many-body Hamiltonian in
\eqref{color_code_Hamiltonian} will present an effective description
of low lying states of the 2-body Hamiltonian \eqref{2-bodyTCC}.
We use the perturbation about the Hamiltonian in \eqref{2-bodyTCC}
considering the coupling regim \eqref{triangle_limit}. Here, the
qubits on the triangles are physical qubits, and logical qubits are
those living at the vertices of the 2-colex. We refer to triangles
as sites, since they correspond to the vertices of the 2-colex. Thus
a triangle will be shown by index $v$ and its vertices by Latin
indices $i,j$. In fact the low lying spectrum of 2-body Hamiltonian
encodes the following projection from the physical qubits to the
logical ones at each site: \be \label{projection}
P_{v}=\ket{\!\Uparrow}\bra{\uparrow\uparrow\uparrow\!}+\ket{\!\Downarrow}\bra{\downarrow\downarrow\downarrow\!},\ee
where $\ket{\!\Uparrow}$ and $\ket{\!\Downarrow}$ stands for the two
states of the logical qubit at site $v$, and $\ket{\!\uparrow}$
($\ket{\!\downarrow}$) is usual up (down) states of a single spin in
computational bases.

To this end, we split the 2-body Hamiltonian into two parts. The
unperturbed part is
$H_{0}=-J_{z}\sum_{\mathrm{b}-link}\sigma^{z}_{i}\sigma^{z}_{j}$. In
the limit of strong Ising interaction the system is polarized. The
interactions between neighboring qubits on different triangles are
included in $V$. They are $\sigma^{x}_{i}\sigma^{x}_{j}$ and
$\sigma^{y}_{i}\sigma^{y}_{j}$ corresponding to red and green links
in Fig.\ref{ruby_lattice}, respectively. So, the transverse part of
the Hamiltonian is \be \label{transverse}
V=-J_{x}\sum_{\mathrm{r}-link}\sigma^{x}_{i}\sigma^{x}_{j}-J_{y}\sum_{\mathrm{g}-link}\sigma^{y}_{i}\sigma^{y}_{j}.\ee
In the case when $J_{z}\gg J_{x}, J_{y}$ the low lying excitations
above the fully polarized state can be treated perturbatively.

The unperturbed part of the Hamiltonian, $H_{0}$, has a highly
degenerate ground space because, for each triangle, two fully
polarized states $\ket{\!\uparrow\uparrow\uparrow}$ and
$\ket{\!\downarrow\downarrow\downarrow}$ have same energy. The
ground state subspace is spanned by all configurations of such
polarized states. Let $N$ be the number of triangles of the lattice.
The ground state energy is $E_{0}^{(0)}=-3NJ_{z}$ and the dimension
of the ground space of the $H_{0}$ or ground state degeneracy reads
$g_{0}=2^{N}$. The first excited state is produced by exciting one
of triangles and has energy $E_{1}^{(0)}=(-3N+4)J_{z}$ with
degeneracy $g_{1}=6N2^{N-1}$. The second excited state has energy
$E_{1}^{(0)}=(-3N+8)J_{z}$ with degeneracy $g_{2}=18N(N-1)2^{N-2}$,
and so on and so forth.

We analyze the effect of $V$ on the ground state manifold by using
the degenerate perturbation theory\cite{Bergman_dpt} in couplings
$J_{x}$ and $J_{y}$. We are interested in how ground state
degeneracy is lifted by including the interaction between triangles
perturbatively. Let $\mathcal{L}$ stand for the ground state
manifold with energy $E^{(0)}_{0}$ and let $P$ be the projection
onto the ground state manifold $\mathcal{L}$. The projection is
obtained from the degenerate ground states as follows: \be
P=\prod_{v\in \Lambda}P_{v} ~~~~~~~~,~~~~~~
P_{v}=|\!\Uparrow\rangle\langle\uparrow\uparrow\uparrow\!|+|\!\Downarrow\rangle\langle\downarrow\downarrow\downarrow\!|.\ee
Using the projection and Green's function we can calculate the
effective Hamiltonian at any order of Perturbation theory. The
eigenvalues of the effective Hamiltonian $H_{\mathrm{eff}}$ appear
as the poles of the Green function $G(E)=P[1/(E-H)]P$. The effect of
perturbation can be recast into the self-energy $\Sigma(E)$ by
expressing the Green's function as $1/(E-E^{(0)}_{0}-\Sigma(E))$.
So, the effective Hamiltonian will be \bea \label{eq4}
H_{\mathrm{eff}}=\sum^{\infty}_{l=0}H^{(l)}_{\mathrm{eff}}=E^{(0)}_{0}+\Sigma(E).\eea
The self-energy can be represented in terms of Feynman diagrams and
can be computed for any order of perturbation: \bea \label{eq5}
\Sigma(E)=PV\sum^{\infty}_{n=0}\mathcal{U}^{n}P,\eea where
$\mathcal{U}=[1/(E-H_{0})](1-P)V$. The energy $E$ can also be
expanded at different orders of perturbation,
$E=E^{(0)}_{0}+\sum^{\infty}_{l=1}E^{(l)}_{0}$. Now, we are at the
position to determine different orders of perturbation. Each term of
$V$ acts on two neighboring physical qubits of different triangles.
At a given order of perturbation theory, there are terms which are
product of $\sigma^{x}$ and $\sigma^{y}$ acting on the ground state
subspace. Each term when acts on the ground state manifold brings
the ground state into an excited state. However, there may be a
specific product of the $\sigma^{x}$ and $\sigma^{y}$ which takes
the ground state into itself, i.e. preserve the polarized
configurations of triangles.

At zeroth-order the effective Hamiltonian will be trivial
$H^{(0)}_{\mathrm{eff}}=E^{(0)}_{0}$. The first-order correction is
given by the operator \bea \label{firstorder}
H^{(1)}_{\mathrm{eff}}=PVP. \eea The effect of $V$ is to move the
states out of the ground state manifold because each term either
$\sigma^{x}\sigma^{x}$ or $\sigma^{y}\sigma^{y}$ flip two qubits
giving rise to two triangles being excited, i.e $VP=P_{2}VP$, where
the operator $P_{2}$ is the projection to second excited state
manifold. Therefore, $PVP=0$, and there is no first-order correction
to the ground state energy.

The second-order correction to the ground state will be the
eigenvalues of the following operator. \bea \label{secondorder}
H^{(2)}_{\mathrm{eff}}=PVG'(E^{(0)}_{0})VP+PVP,\eea where the
operator $G'(E)=1/(E-H_{0})$ is the unperturbed Green's function and
the superscript prime stands for the fact that its value be
\emph{zero} when acts on the ground state. The second-order
correction only shifts the ground state energy, and therefore, the
second-order effective Hamiltonian acts trivially on ground state
manifold,\bea \label{effective2}
H^{(2)}_{\mathrm{eff}}=3N\frac{J^{2}_{x}+J^{2}_{y}}{-8J_{z}}P.\eea
In fact the first $V$ flips the qubits and the second $V$ flips them
back. As we go to higher order of perturbation theory the terms
become more and more complicated. However, if the first-order is
zero as in our case, the terms becomes simpler. Thus, the
third-order of perturbation will be zero and will leave corrections
to energy and ground state intact: \bea
\label{thirdorder}H^{(3)}_{\mathrm{eff}}=PV \left( G'(E^{(0)}_{0})V
\right)^{2}P=0. \eea The forth-order of perturbation theory
contributes the following expression to the correction of ground
state manifold:\bea \nonumber H^{(4)}_{\mathrm{eff}}=PV\left(
G'(E^{(0)}_{0})V
\right)^{3}P-E^{(2)}_{0}PV\left( G'(E^{(0)}_{0})\right)^{2}VP,\\
\label{forthorder}\eea where $E^{(2)}_{0}$ is the second order
correction to the ground state energy obtained in
\eqref{effective2}. The first term includes four $V$ and must act in
the ground state in which the last $V$ returns the state to the
ground state manifold. The second term is like the second-order.
There are many terms which must be calculated. However, since the
forth-order only gives a shift to the ground state energy, we don't
need them explicitly. So, we can skip the forth-order. Fifth-order
correction yields terms each containing odd number of $V$, so it
gives zero contribution to the effective Hamiltonian.

The sixth-order of perturbation leads to the following long
expression.
 \bea \nonumber H^{(6)}_{\mathrm{eff}}&=&PV\left(
G'(E^{(0)}_{0})V \right)^{5}P-E^{(4)}_{0}PV\left(
G'(E^{(0)}_{0})\right)^{2}VP\\ \nonumber
&&+\left(E^{(2)}_{0}\right)^{2}PV\left( G'(E^{(0)}_{0})\right)^{3}VP
\\ \nonumber &&
-E_{0}^{(2)}PV\left(G'(E^{(0)}_{0})\right)^{2}V\left(G'(E^{(0)}_{0})V
\right)^{2}P\\ \nonumber &&-E_{0}^{(2)}PV G'(E^{(0)}_{0})V\left(
G'(E^{(0)}_{0})\right)^{2}V G'(E^{(0)}_{0})VP\\
\label{sixthorder} &&-E_{0}^{(2)}PV\left( G'(E^{(0)}_{0})V
\right)^{2}\left( G'(E^{(0)}_{0})\right)^{2}VP.  \eea
Apart from the first term, other terms contain two or four $V$ and
as we discussed in the preceding paragraphs they only contribute a
shift in the ground state energy. However, the first term gives the
first non-trivial term breaking in part the ground state degeneracy.
In the sixth order correction, there are some terms which are the
product of $\sigma^{x}\sigma^{x}$ and $\sigma^{y}\sigma^{y}$
associated to the red and green links of the ruby lattice. Some
particular terms, as seen below, may map ground state subspace into
itself. For instance, consider the following product of links around
an inner hexagon \bea \label{z_plaquette}
\prod_{<i,j>}\sigma^{w}_{i}\sigma^{w}_{j}=\pm\prod_{i\in
V_{h}}\sigma_{i}^{z}, \eea where the first product runs over three
red and three green links making an inner hexagon, $V_{h}$ stands
for the set of its vertices and the prefactor $\pm$ depends on the
ordering of links in the product. The action of a $\sigma^{z}$ on
one vertex (qubit) of a triangle encodes an logical $\tau_{v}^{z}$
operator acting on the associated vertex of lattice $\Lambda$. This
can explicitly be seen from the following relation: \bea
\label{zencoded}
\tau_{v}^{z}=P_{v}\sigma^{z}P_{v}=|\!\Uparrow\rangle\langle\Uparrow\!|-|\!\Downarrow\rangle\langle\Downarrow\!|,\eea
where $\sigma^{z}$ acts on one of the vertices of a triangle and
$P_{v}$ is the projection defined in \eqref{projection}. Thus, the
expression of \eqref{z_plaquette} can be related to the plaquette
operator $B^{z}_{f}=-\prod\tau_{v}^{z}$, where the index $f$ denotes
a plaquette of effective lattice $\Lambda$ as in
Fig.\ref{color_code} and product runs over six sites around it. Now
we go on to pick up the sixth-order correction to the ground state
manifold. There are many terms which must be summed. Sixth-order
correction up to a numerical constant contributes the following
expression to the effective Hamiltonian:\bea \label{effective6}
H^{(6)}_{\mathrm{eff}}=\mathrm{constant}-\delta\frac{J_{x}^{3}J_{y}^{3}}{J_{z}^{5}}\sum_{f}B^{z}_{f},
\eea  where $\delta$ is a positive numerical constant arising from
summing up $720$ terms related to the order of product of six links
around an inner hexagon of ruby lattice. Although, its exact
numerical value is not important, but knowing its sign is essential
for our subsequent discussions. As it is clear from the first term
in \eqref{sixthorder}, five Green's functions and six $V$ in the
perturbation contribute a minus sign to the expression. This minus
sign together with the sign appearing in \eqref{z_plaquette} enforce
the coefficient $\delta$ be a positive constant. Now it is simple to
realize how the vectors in the ground state manifold rearranged.
Trivially, all plaquette operators $B^{z}_{f}$ commute with each
other and their eigenvalues are $\pm1$. All polarized vectors in
$\mathcal{L}$ are the eigenvector of the effective Hamiltonian
emerging at sixth-order. But, those are ground states of the
effective Hamiltonian in \eqref{effective6} which are eigenvectors
of all plaquette operators $B^{z}_{f}$ with eigenvalue $+1$. Thus,
highly degenerate ground state of the unperturbed Hamiltonian is
broken in part. The same plaquette operators $B^{z}_{f}$ also appear
at higher order of perturbation. For example, at eighth-order.
Instead of giving the rather lengthy expression of the eighth-order
correction, we only keep terms resulting in the plaquette operators
as follows: \bea \label{effective8}
H^{(8)}_{\mathrm{eff}}=\mathrm{constant}-
\beta\frac{J_{x}^{5}J_{y}^{3}+J_{x}^{3}J_{y}^{5}}{J_{z}^{7}}\sum_{f}B^{z}_{f},\eea
where $\beta>0$. This term is added to the one in \eqref{effective6}
to give the effective Hamiltonian up to eighth-order, but the ground
state structure remains unchanged. Further splitting in the ground
state manifold is achieved by taking into account the ninth-order of
perturbation. The expression of ninth-order is very lengthy.
However, the first term of the expression containing nine $V$ gives
some terms being able of mapping the ground state manifold into
itself in a nontrivial way. These terms map a polarized triangle,
say up, to a down one. Indeed, when one or two qubits of the
polarized triangle gets flipped, its state is excited. However,
flipping three qubits of the triangle returns back the ground state
onto itself. This process encodes $\tau^{x}$ and $\tau^{y}$ logical
operators acting on logical qubits arising through the projection.
Let $\sigma^{x}_{1}$, $\sigma^{x}_{2}$ and $\sigma^{y}_{3}$ act on
three qubits of a triangle. The encoded $\tau^{y}$ operator will be
\bea \label{yencoded}
\tau_{v}^{y}=P_{v}\sigma^{x}_{1}\sigma^{x}_{2}\sigma^{y}_{3}P_{v}=-i|\!\Uparrow\rangle\langle\Downarrow\!|+i|\!\Downarrow\rangle\langle\Uparrow\!|.\eea
If $\sigma^{x}_{1}$, $\sigma^{y}_{2}$ and $\sigma^{y}_{3}$ act on
three qubits of a triangle, the encoded $\tau^{x}$ logical operator
will be \bea \label{xencoded}
-\tau_{v}^{x}=P_{v}\sigma^{x}_{1}\sigma^{y}_{2}\sigma^{y}_{3}P_{v}=-|\!\Uparrow\rangle\langle\Downarrow\!|-|\!\Downarrow\rangle\langle\Uparrow\!|.\eea
As we already pointed out a plaquette of the ruby lattice is made up
of an inner hexagon, an outer hexagon and the six blue triangles. It
is possible to act on the polarized space of the blue triangles by
making two different combinations of $9$ link interactions: i/
Applying $6$ link interactions on the outer hexagon (three of them
of XX type and another three of YY type), times 3 link interactions
of red type on the inner hexagon. Notice that every vertex of the
blue triangles in the plaquette gets acted upon these $9$ link
interactions. The resulting effective operator is of type $\tau^y$
due to \eqref{yencoded}.  ii/ Applying $6$ link interactions on the
outer hexagon, times $3$ link interactions of green type. Then, the
resulting effective operator is of type $\tau^x$ due to
\eqref{xencoded}.

The effective Hamiltonian at this order then reads \bea
H^{(9)}_{\mathrm{eff}}&=&PV\left( G'(E^{(0)}_{0})V
\right)^{8}P+...\\ \nonumber &=&
\mathrm{constant}-\gamma\frac{J_{x}^{6}J_{y}^{3}}{J_{z}^{8}}\sum_{f}B^{y}_{f}-\gamma\frac{J_{x}^{3}J_{y}^{6}}{J_{z}^{8}}\sum_{f}B^{x}_{f}\label{effective9}.\eea
Again, the sign of coefficient $\gamma$ is important. Nine $V$'s,
six $\tau^{x}$ or $\tau^{y}$, and eight Green's function imply that
the $\gamma$ must have positive sign.

Putting together all above obtained corrections lead to an effective
Hamiltonian encoding color code as its ground
state\cite{topodistill, entanglementTCC}. Therefore, up to constant
terms, the effective Hamiltonian reads as follows \bea
\label{effective_Hamiltonian}
H_{\mathrm{eff}}=-k_{z}\sum_{f}B^{z}_{f}-k_{x}\sum_{f}B^{x}_{f}-k_{y}\sum_{f}B^{y}_{f},\eea
where the $k_{z}$, $k_{x}$ and $k_{y}$ are positive coefficients
arising at different orders. Since $B^{x}_{f}B^{y}_{f}=B^{z}_{f}$,
the above effective Hamiltonian is just the many-body Hamiltonian of
the color code as in \eqref{stabilizer_Hamiltonian}. The terms
appearing in the Hamiltonian mutually commute, so the ground state
will be the common eigenvector of plaquette operators. Since each
plaquette operator squares identity, the ground state subspace,
$\mathcal{C}$, spanned by vectors which are common eigenvectors of
all plaquette operators with eigenvalue $+1$, i.e \bea
\label{codingspace} \mathcal{C}=\{|\psi\rangle:~
B^{x}_{f}|\psi\rangle=|\psi\rangle,~
B^{y}_{f}|\psi\rangle=|\psi\rangle~;~~\forall f \}.\eea

The group of commuting boundary closed string operators can be used
as an alternative way to find the terms appearing in the effective
Hamiltonian\cite{Kells}. As we pointed out in the preceding section,
the non-zero contribution from various orders of perturbation theory
results from the product of red and green links which preserve the
configurations of the polarized triangles, i.e maps ground state
manifold onto itself. For instance consider the elementary plaquette
operator $A$ corresponding to a closed string in
Fig.\ref{plaquette_IOM_Independent}. Each triangle contributes
$\sigma^{y}\sigma^{y}\sigma^{x}$ to the expression of operator which
is projected to $\tau^{x}$ as in \eqref{xencoded}. Thus, the
effective representation of plaquette operator reads as follows:
\bea \label{plaquette_projectionA} PS_{f}^{A}P \rightarrow
B^{x}_{f}.\eea Plaquette operators $S_{f}^{B}$ and $S_{f}^{C}$ can
also be recast into the effective forms as follows: \bea
\label{plaquette_projectionB} PS_{f}^{B}P \rightarrow
B^{y}_{f}~,~~~~~~~~~~~PS_{f}^{C}P \rightarrow -B^{z}_{f}.\eea These
are lowest order contributions to the effective Hamiltonian as we
obtained in \eqref{effective_Hamiltonian}. Higher order of
perturbation will be just the product of effective plaquette
operators. The nontrivial strings winding around the torus will have
also effective representations and appear at higher orders of
perturbation. In general, every string-net IOM on the ruby lattice
are projected on an effective one as in
Fig.\ref{stringnet_effective}.

\section{Bosonic Mapping}
\label{sect_V}
As we stated in Sect.\ref{sect_III},  one of the defining properties
of our model is the existence of non-trivial integrals of motion
IOM, called string-nets. As a particular example, the Kitaev's model
on the honeycomb lattice has strings IOMs, but not string-nets. We
are interested in models in which the number of IOMs is proportional
to the number of qubits in the lattice, i.e., \be I = \eta N_q, \ee
where $I$ is the number of IOMs, $N_q$ is the number of qubits
(spins) in the lattice, and $\eta$ is a fraction: $\eta=\half$ for
the Kitaev model \cite{honeycomb}, $\eta=\frac{1}{3}$ for our model
in \eqref{2-bodyTCC} \cite{AnyonicFermions}. The fact that $\eta$ is
a fraction $\eta \leq1$ implies that these models based on
string-net IOMs will not necessarily be exactly solvable. In the
Kitaev's model, it turns out to be exactly solvable using an
additional mapping with Majorana fermions, but this need not be a
generic case. Therefore, we need to resort to other techniques in
order to study the physical properties of these models. We consider
here and next section, approximate methods based on bosonic and
fermionic mappings. The application of the bosonic method to our
model is based on a mapping from the original spins on the blue
triangles to hardcore bosons with spin \cite{AnyonicFermions}. With
this mapping, it is possible to use the PCUTs approach (Perturbative
Continuous Unitary Transformations) \cite{pcut_Uhrig}. This is
inspired by the RG method based on unitary transformation introduced
by Wegner (the Similarity RG method) \cite{pcut_Wegner}. Originally,
the PCUTs method was applied to the Kitaev model
\cite{boson_honeycomb}. As we will see below, the method paves the
way to go beyond the perturbation approach presented in the
preceding section which fits into a sector without any hard-core
boson. The physics at other sectors is very promising and we study
it here. Emergence of three families of strongly interacting anyonic
fermions and invisible charges are among them, which are not present
in the Kitaev's model or its variants. To start with, let us set for
simplicity $J_{z}=1/4$, and consider the extreme case of
$J_{x}=J_{y}=0$. In this case the system consists of isolated
triangles. The ground states of an isolated triangle are polarized
states $\ket{\!\uparrow\uparrow\uparrow}$ and
$\ket{\!\downarrow\downarrow\downarrow}$ with energy $-3/4$. The
excited states that appear by flipping spins are degenerate with
energy $1/4$. In this limit, the spectrum of whole system is made of
equidistance levels being well-suited for perturbative analysis of
the spectrum: Green's function formalism as discussed in the
preceding section, which may capture only the lowest orders of
perturbation or another alternative approach based on the PCUT. The
change from the ground state to an exited state can be interpreted
as a creation of particles with energy $+1$. This suggest an exact
mapping from the original spin degrees of freedom to quasiparticles
attached to effective spins. The mapping is exact, i.e. we don't
miss any degrees of freedom. Such a particle is a hard-core boson.
At each site, we attach such a boson and also an effective
spin-$\frac{1}{2}$. Let choose the following bases for the new
degrees of freedom \be |a,d\rangle=|a\rangle\otimes|d\rangle,
~~~~~~a=\Uparrow,\Downarrow,~~~~d=0,\mathrm{r,g,b},
\label{spinboson_bases}\ee where $a$ and $d$ stand for states of the
effective spin and quasiparticle attached to it. The Hilbert space
$\mathcal{H}_{C}$ representing the hard-core bosons is four
dimensional spanned by bases
$\{\ket{0},\ket{\mathrm{r}},\ket{\mathrm{g}},\ket{\mathrm{b}}\}$.
Now the following construction relates the original spin degrees of
freedom and new ones in \eqref{spinboson_bases} \bea \nonumber
\ket{\!\Uparrow,0}\equiv\ket{\!\uparrow\uparrow\uparrow},~~~~~&&\ket{\!\Downarrow,0}\equiv\ket{\!\downarrow\downarrow\downarrow}\\
\nonumber \ket{\!\Uparrow,\mathrm{r}}\equiv\ket{\!\uparrow\downarrow\downarrow},~~~~~&&\ket{\!\Downarrow,\mathrm{r}}\equiv\ket{\!\downarrow\uparrow\uparrow} \\
\nonumber \ket{\!\Uparrow,\mathrm{g}}\equiv\ket{\!\downarrow\uparrow\downarrow},~~~~~&&\ket{\!\Downarrow,\mathrm{g}}\equiv\ket{\!\uparrow\downarrow\uparrow}\\
\ket{\!\Uparrow,\mathrm{b}}\equiv\ket{\!\downarrow\downarrow\uparrow},
~~~~~&&\ket{\!\Downarrow,\mathrm{b}}\equiv\ket{\!\uparrow\uparrow\downarrow}.
\label{mapping_bases} \eea Within such mapping, the effective spins
and hard-core bosons live at the sites of the effective hexagonal
lattice $\Lambda$ in Fig.~\ref{lattice_letter}(c). Recall that this
lattice is produced by shrinking the triangles. At each site we can
introduce the color annihilation operator as $b_{c} := \ket 0 \bra
c$. The number operator $n$ and color number operator $n_{c}$ are
\be n :=\sum_c n_c, \qquad n_c := b^\dagger_c b_c. \ee Annihilation
and creation operators anticommute on a single site, and commute at
different sites, that is why they are hard-core bosons. We can also
label the Pauli operators of original spins regarding to the their
color in Fig.~\ref{lattice_letter}(b) as $\sigma^{w}_{c}$ with
$c=\mathrm{r,g,b}$. The mapping in \eqref{mapping_bases} can be
expressed in operator form as follows
\begin{equation}\label{mapping} \sigma^z_c \equiv \tau^z \otimes
p_c, \quad \sigma^\nu_c \equiv \tau^\nu \otimes (b_c^\dagger + b_c +
s_\nu r_c), \end{equation} where $\nu=x,y$, $s_x:=-s_y:=1$, the
symbols $\tau$ denote the Pauli operators on the effective spin and
we are using the color parity operators $p_c$ and the color
switching operators $r_c$ defined as
\begin{equation}\label{parity}
p_c := 1-2(n_{\bar c}+n_{\bbar c}),\qquad r_c := b^\dagger_\bc
b_\bbc + b^\dagger_\bbc b_\bc.\end{equation} Now we can forget the
original ruby lattice and work on the effective lattice in which the
bosons are living at its sites. With the above identification for
Pauli operators, the 2-body Hamiltonian in \eqref{2-bodyTCC} can be
written in this language. Before that, let fix a simplified
notation. All spin and bosonic operators act on the sites of the
effective lattice. We refer to a site by considering its position
relative to a reference site: the notation $O_{\rel c}$ means $O$
applied at the site that is connected to a site of reference by a
$c$-link. The 2-body Hamiltonian then becomes
\begin{equation}\label{Hamiltonian_B}
H=-3N/4+Q-\sum_{\Lambda}\sum_{c\neq c\prima} J_{c\prima|c}\,
T_c^{c\prima},\end{equation} with $N$ the number of sites,
$Q:=\sum_\Lambda n$ the total number of hardcore bosons,  the first
sum running over the $N$ sites of the reduced lattice, the second
sum running over the 6 combinations of different colors $c,c\prima$
and\begin{equation}\label{Terms}
 T_c^{c\prima} = u_c^{c\prima}+ \frac {t_c^{c\prima}+v_c^{c\prima}} 2 + \frac {r_c^{c\prima}} 4 + \mathrm{h.c.},
\end{equation}
a sum of several terms for an implicit reference site, according to
the notation convention we are using. The meaning of the different
terms in \eqref{Terms} is the following. The operator
$t_c^{c\prima}$ is a $c$-boson hopping, $r_c^{c\prima}$ switches the
color of two $\bc$- or $\bbc$-bosons, $u_c^{c\prima} $ fuses a
$c$-boson with a $\bc$-boson (or a $\bbc$-boson) to give a
$\bbc$-boson ($\bc$-boson) and $v_c^{c\prima}$ destroys a pair of
$c$-bosons. The explicit expressions are
\begin{alignat}{2} \no
&t_c^{c\prima} := \tau_c^{c\prima} b_c b^\dagger_{c\rel
c\prima},\qquad &r_c^{c\prima} := \tau_c^{c\prima} r_c r_{c\rel
c\prima},\\ &u_c^{c\prima} :=  s_{c\prima|c} \tau_c^{c\prima} b_c
r_{c\rel c\prima}, \qquad &v_c^{c\prima} := \tau_c^{c\prima} b_c
b_{c\rel c\prima}, \label{terms} \end{alignat} where we are using
the notation
\begin{equation}
\tau_c^{c\prima}:=\tau^{c\prima|c}\tau_{\rel c\prima}^{c\prima|c}.
\end{equation}
We can also describe the plaquette IOM operators in
Fig.~\ref{plaquette_IOM_Independent} in terms of spin-boson degrees
of freedom by means of the mapping in \eqref{mapping}. For each
plaquette $f$ and color $c$, the plaquette operator is expressed as
\begin{equation} \label{Plaquette_boson} S_f^c := \prod_{v\in f} \tau_{v}^{c|c\prima}
p_{c\prima\star c},
\end{equation} where $c\prima$ is the color of the plaquette $f$,
the product runs through its sites and $\star$ is just a convenient
symmetric color operator defined by
\begin{equation}
c\star c:=c,\qquad c\star \bc:= \bc\star c:=\bbc.
\end{equation}
The relation in \eqref{Plaquette_boson} is just a generalization of
plaquette operators in \eqref{plaquette_honycomb1} to other sectors
of the system. In fact taking the zero particle sector, the
expressions in \eqref{plaquette_honycomb1} are recovered. In the
same way the nontrivial string operators in Fig.\ref{strings_IOM}
can be described with the above mapping as
\begin{equation} S^c_\mu:=\prod_{v\in \mu} \tau_{v}^{c|c\prima}p_{c\star
c\prima},\end{equation} where $\mu$ denotes the homology class of
the string. On closed surfaces, not all plaquette operators are
independent. They are subject to the following constraints
\begin{equation}\label{constraints} \prod_{f\in\Lambda} S_{f}^c
= (-1)^{N/2}, \qquad \prod_{c=\mathbf{r, g, b}} S_{f}^c =
(-1)^{s/2},\end{equation} where $s$ is the number of sites of a
given plaquette $f$. The first equation can be further divided into
products over subsets of plaquettes giving rise to other constants
of motion, the so called color charges, as
\begin{equation}\label{constraints_charge} \prod_{f\in\Lambda_{\bc}} S_{f}^\bc\prod_{f\in\Lambda_{\bbc}}
S_{f}^\bbc=\prod_{f\in\Lambda_{\bbc}}
S_{f}^{c}\prod_{f\in\Lambda_{c}} S_{f}^\bc=
\prod_{\Lambda}p_{c}.\end{equation} In these products the spin
degrees of freedom are washed out, since they appear twice and
consequently square identity. By use of equation \eqref{parity}, the
product over parity operators can be written as \be
\label{parity_charge}
\chi_\Lambda(c)=\prod_{\Lambda}p_{c}=(-1)^{Q_{\bc}+Q_{\bbc}},\ee
where $Q_{c}=\sum_{\Lambda}n_{c}$ is the total number of $c$-bosons.
It is simple to check that the above equation commutes with
Hamiltonian in \eqref{Hamiltonian_B}. For each family of bosons we
can attach a charge. We suppose that each $c$-boson carries a charge
as $\chi_\mathrm{c}$, that is, an irrep of the gauge group. In
particular, the Hamiltonian preserves the following total charge
\begin{equation}\label{color_charge}
\chi_\Lambda:=\chi_\mathrm{r}^{Q_\mathrm{r}}\chi_\mathrm{g}^{Q_\mathrm{g}}\chi_\mathrm{b}^{Q_\mathrm{b}}.
\end{equation}
\subsection{Emerging particles: anyonic fermions}

Equation \eqref{constraints_charge} could already suggest that the
parity of vortices are correlated to the parity of the number of
bosons. In particular, creation of a $c$-boson changes the vorticity
content of the model.

\begin{figure}
\begin{center}
\includegraphics[width=12cm]{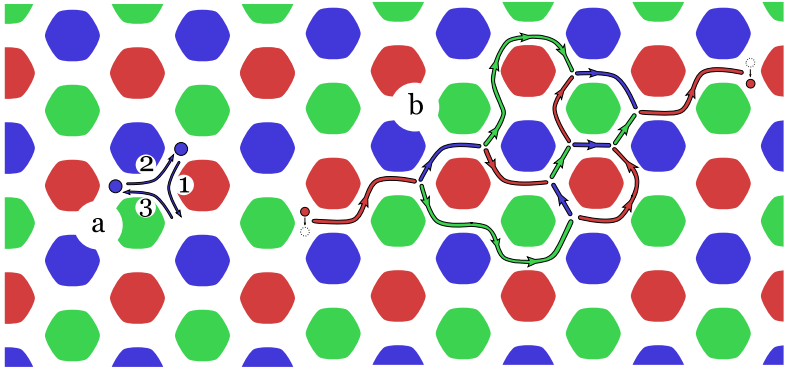} \caption{(color online) (a)A step by step
illustration of hopping of two blue fermions with the net result of
exchange of two fermions (b)A representation of a fermionic process
in which a $c$-fermion (here red) at the origin site is annihilated
and undergoes hopping, splitting and fusion processes, and then is
created at another site.} \label{fermionic_process}
\end{center}
\end{figure}

The statistics of vortices depend on their color and type as in
Fig.\ref{topological_charges}. But what about the statistics of
$c$-bosons? As studied in \cite{wen_statistics}, the statistics of
quasiparticles can be examined using the hopping terms. These
hopping terms are combined so that two quasiparticles are exchanged.
In addition to usual hopping terms, we also need for composite
hopping , that is, a $c$-boson hops from a $c$-plaquette to another,
which is carried out by terms like $t_c^c=u_\bbc^c
{u_{\bc,c}^{c\,\dagger}}=u_\bc^c {u_{\bbc,c}^{c\,\dagger}}$. Let us
consider a state with two $c$-boson excitations located at two
different sites separated from a reference site by $\bc$ and $\bbc$
links. An illustrative example for the case of, say, blue bosons is
depicted in Fig.\ref{fermionic_process}(a). Consider a process with
the net effect of resulting into the exchange of two bosons. Each
step of process can be described by hopping terms. Upon the
combination of hopping terms, we are left with the following phase
\begin{equation}\label{fermions}
 t_{c}^\bbc \,t_{c,c}^c \, t_{c}^\bc \,  t_{c,\bbc}^\bbc \, t_c^c \, t_{c,\bc}^\bc   =
(\tau^y\tau^y_{,\bbc}\tau^z\tau^z_{,c}\tau^x\tau^x_{,\bc})^2=-1,
\end{equation}
which explicitly show that the quasiparticles made of
hardcore-bosons and effective spins have fermion statistics
\cite{AnyonicFermions}. Thus we have three families of fermions each
of one color. These are high energy fermions interact strongly with
each other due to the fusion term in the Hamiltonian. Fermions from
different families have mutual semionic statistics, that is,
encircling one $c$-fermion around a $\bc$-fermion picks up a minus
sign. This can also be checked by examining the hopping terms as in
Fig.\ref{fermionic_process}(a). Thus we are not only dealing with
fermions but also with anyons.

The elementary operators in \eqref{terms} have a remarkable
property, and that, they all commute will plaquettes and strings
IOM. This naturally implies any fermionic process leaves the
vorticity content of the model unaffected. A fermionic process may
correspond to hopping, splitting, fusion and annihilation driven by
the terms in the Hamiltonian \eqref{Terms} and \eqref{terms}. A
typical fermionic process is shown in
Fig.\ref{fermionic_process}(b), with the net result of displacement
of a r-fermion from a site at the origin to other one of the
effective lattice. A very feature of this process is the existence
of vertexes, which is essential to high-energy fermions, that is,
three fermions with different color charge can fuse into the vacuum
sector. At the vertex, three different colored strings meet. Notice
that the colored strings shown here have nothing to do with the ones
we introduced in Sect.\ref{sect_III}. Indeed, these are just product
of some red and green links of the ruby lattice. When translated
into the spin-boson language, they are responsible for
transportation of $c$-fermions through the lattice.

Now we can think of constraints in \eqref{constraints_charge} and
\eqref{parity_charge} as a correlation between the low energy and
high energy sectors of our model. They explicitly imply that the
creation of a $c$-fermion creates the vortices with net topological
charge of $(\bc,\chi_{\bbc})$. Alternatively, as suggested by
mapping in \eqref{mapping_bases}, flipping of spin on a triangular
can create or destroy the excitation, that is, a high energy fermion
can be locally transformed into low energy ones. This amounts to
attach a topological charge from low energy sector to high energy
excitations. Thus a $c$-fermion carries a topological charge. On the
other hand, an open $c$-string commutes with all plaquettes except
some of them, so they create or destroy a particular charge among
the charges in Fig.\ref{topological_charges}. It is simple to check
that which charge they carry at their open ends. In fact they carry
$(\bbc,\chi_{\bc})$ charges. These latter charges have trivial
mutual statistics relative to the charges carried by high energy
fermions, since they belong to different families of fermions. Thus
the charge carried by a $c$-string must be invisible to the high
energy fermions.
\subsection{Perturbative continuous unitary transformation}
The physics behind Hamiltonian in \eqref{Hamiltonian_B} can be
further explored by resorting to approximate methods. The specific
form of the energy levels of our model in the isolated limit, the
existence of equidistant levels, makes it suitable for perturbative
continuous unitary transformations. In this method the Hamiltonian
is replaced by an effective one within unitary transformations in
which the resulting effective Hamiltonian preserves the total
charges, i.e. $[H^{\mathrm{eff}},Q]=0$. Thus the analysis of the
model relies on finding the effective Hamiltonian in a sector
characterized by the number of charges at every order of
perturbation. For our model each sector are determined by the number
of $c$-fermions. Each term of the effective Hamiltonian in any
sector is just a suitable combination of expressions in
\eqref{terms} in such a way that respects the total color charge in
that sector. For now we briefly analyze the lowest charge sectors.

In the zero-charge sector, only the effective spin degrees of
freedom do matter. The effective Hamiltonian is just a many-body
Hamiltonian with terms that are product of plaquette operators as
follows
\begin{equation} \label{0_sector_Hamiltonian}
H_{0}^{\mathrm{eff}}=E_{0}-\sum_{\{c\}}\sum_{\{f\}}O^{c_{1},...,c_{n}}_{f_{1},...,f_{n}}S^{c_{1}}_{f_{1}}...S^{c_{n}}_{f_{n}},
\end{equation}
where the first and second sum run over an arbitrary collection of
colors and plaquettes of effective honeycomb lattice. The
coefficients $O$'s are determined at a given order of perturbation.
The product of plaquette operators is nothing but the string-net
operators. Let us focus at lowest order of perturbation, where the
model represents non-interacting vortices. First, let us redefine
plaquette operators as
\begin{equation}
B_f^x=j_x^{s/2}\prod_{v\in f} \tau_{v}^x,\qquad
B_f^y=j_y^{s/2}\prod_{v\in f} \tau_{v}^y,
\end{equation}
where $j_w:=J_w/|J_w|$. At ninth order of perturbation the effective
Hamiltonian is
\begin{equation}\label{Hamiltonian_eff_0}
 H^{\mathrm{eff}}_{0} = - \sum_{f\in\Lambda} \left ( k_x B^x_f + k_yB^y_f + k_z B^x_f B^y_f \right),
\end{equation}
with \cite{AnyonicFermions}
\begin{equation}\label{coeff}
k_{z}=\frac{3}{8}|J_xJ_y|^{3}+O(J^{7}),~~~
\frac{k_{x}}{|J_{y}|^{3}}=\frac{k_{y}}{|J_{x}|^{3}}=\frac{55489}{13824}{|J_{x}J_{y}|^{3}}.
\end{equation}
This is exactly the many-body Hamiltonian of topological color code
obtained in \eqref{effective_Hamiltonian} using degenerate
perturbation theory with the additional advantage of knowing the
coefficients exactly. Its ground state is vortex free and can be
written explicitly by choosing a reference state as
\begin{equation}\label{groundstate} 2^{N/4-1}\prod_{f}\left(\frac{1+B^x_f}{2}\right)\ket{\Uparrow}\otimes\ket{0}_{b}.
\end{equation}
Other degenerate ground states can be constructed by considering the
nontrivial string operators winding around the torus. Excitations
above ground state don't interact. Going to higher order of
perturbation, as equation in \eqref{0_sector_Hamiltonian} suggests,
the ground states remain unchanged, however the excitation spectrum
changes and vortices interact with each other.

The one-quasiparticle sector can also be treated by examining the
expressions in \eqref{terms}. The effective Hamiltonian can be
written as \begin{equation}\label{1_sector_Hamiltonian}
H^{\mathrm{eff}}_{1}=H^{\mathrm{eff}}_{0}-\sum_{\{R\}}O_{R}\hat{R}b^{\dagger}_{c,R}b_{c}.
\end{equation}
What the second term describes is nothing but the annihilation of a
$c$-fermion at a reference site and then its creation at a site
connected to the reference by a string-net $R$, as shown in
Fig.\ref{fermionic_process}(b). Again notice that this string-net is
just the product of green and red links of original 2-body
Hamiltonian, in its effective form is given in terms of spin-boson
degrees of freedom. The coefficients $O$'s are determined at any
order of perturbation. Notice that these coefficients are different
from those in \eqref{0_sector_Hamiltonian}. In the first order, only
the hopping term does matter. Let us consider the sector containing
a $c$-fermion. Up to this order, the fermion can only hop around a
$c$-plaquette. This implies that at first order the fermion perform
an orbital motion around a plaquette of its color. Notice that the
fermion can not hop from a $c$-plaquette to other $c$-plaquette at
the first order, since it needs for a composite process which
appears at second order. This composite process is a combination of
splitting and fusion processes. This is a virtual process in the
sense that the splitting of a $c$-fermion into two $\bc$- and
$\bbc$-fermion takes the model from 1-quasiparticle sector into the
2-quasiparticle sector, but the subsequent process fuses two
particles into a single one turning back to 1-quasiparticle sector.
Thus, at second order the $c$-fermion can jump from one orbit to
other one.

At first order, for $J=J_x=J_y$, we get a $-2J$ contribution to the
energy gap coming from orbital motion. Going to second order we get
a non-flat dispersion relation. The gap, at this order, is given by
$1-2J-J^2/2$ and thus it closes at $J\simeq 0.45$. This is just an
approximate estimation, since we are omitting all fermion
interactions and, perhaps more importantly, we are taking $J\simeq
J_z$. However, it is to be expected that as the couplings $J_x\sim
J_y$ grow in magnitude the gap for high-energy fermions will reduce,
producing a phase transition when the gap closes. Such a phase
transition resembles the anyon condensations discussed in
\cite{family_non_Abelian,nestedTO,bais_slingerland09}. There are
three topological charges invisible to the condensed anyons. This
means that in the new phase there exists a residual topological
order related to these charges. They have semionic mutual statistics
underlying the topological degeneracy in the new phase.

\subsection{Fermions and gauge fields}
The emerging high-energy $c$-fermions always appear with some
nontrivial gauge fields\cite{AnyonicFermions,wen_statistics}, and
carry different representation of the gauge symmetry
$\mathbf{Z}_2\times \mathbf{Z}_2$ of the model. Before clarifying
this, we can see that the plaquette degrees of freedom correspond to
$\mathbf{Z}_2\times \mathbf{Z}_2$ gauge fields. This correspondence
is established via introducing the following plaquette
operators\begin{align}\label{Bs}
B_f^\bbc &:=j_x^{s/2}\, S_f^\bc,\nonumber\\
B_f^\bc &:=j_y^{s/2} \,S_f^\bbc,\nonumber\\
B_f^c &:=(-j_xj_y)^{s/2}\,  S_f^c.
\end{align} The gauge element $q_f\in \mathbf{Z}_2\times \mathbf{Z}_2$ that can
be attached to the plaquette $f$ is determined by following
eigenvalue conditions \begin{align}\label{gpi} \chi_c(q_f) =
B_f^c,\end{align}which always has a solution due to \be \label{BBB}
(B_f^c)^2=B^{\mathrm{r}}_{f} B^{\mathrm{g}}_f B^{\mathrm{b}}_{f}=
1.\ee
\begin{figure}
\begin{center}
\includegraphics[width=11cm]{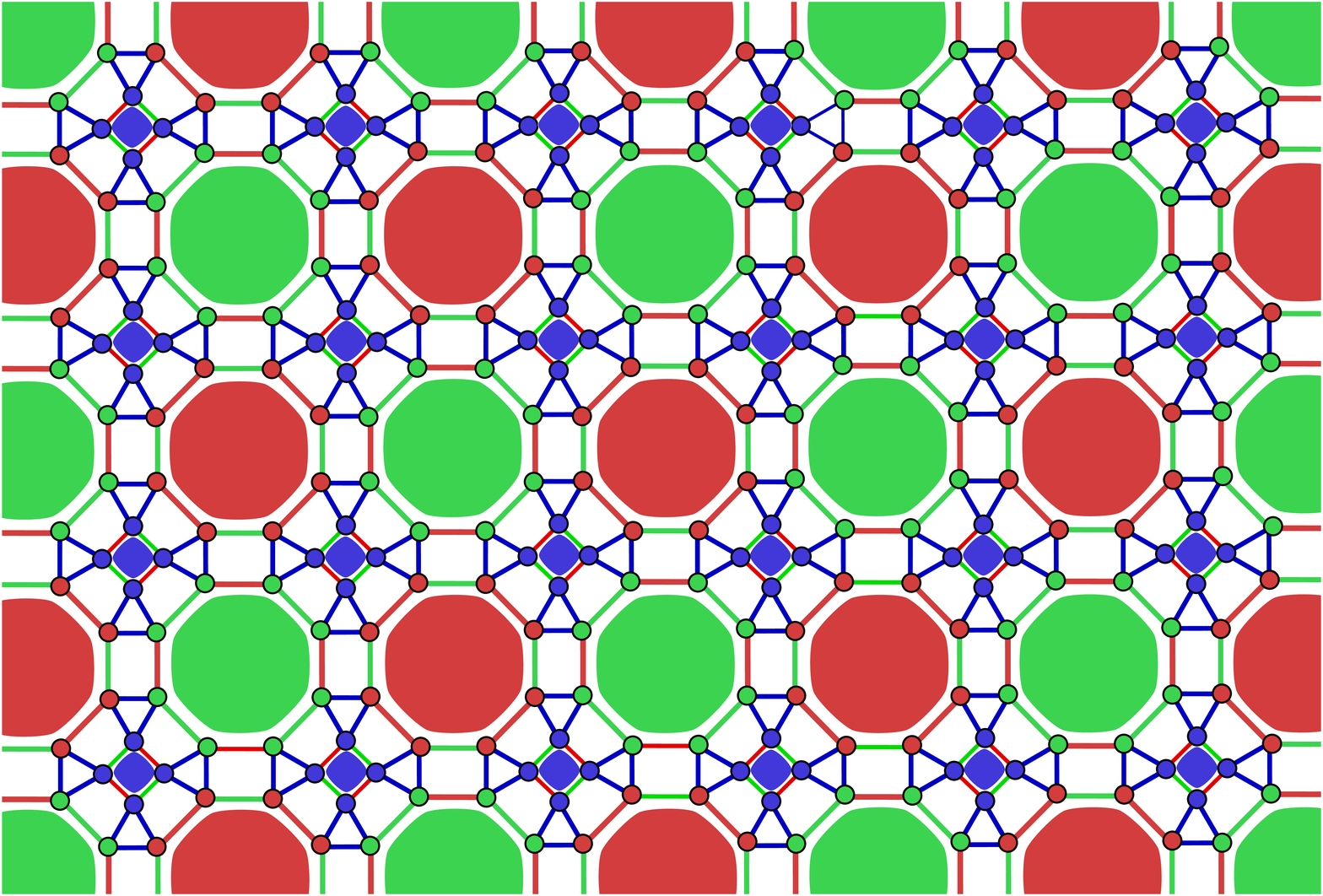} \caption{(color online) A piece
of 2-body lattice corresponding to a 2-colex composed of octagon and
square plaquettes. The lattice is 3-colorable as in
Fig.\ref{lattice_letter}.} \label{foureight}
\end{center}
\end{figure}
The ground state of color code Hamiltonian \eqref{Hamiltonian_eff_0}
is vortex free and corresponds to $\chi_c=1$. The fact that for a
2-colex with hexagonal plaquettes, the gauge fields can be related
to representation of the group is immediate. One way to see this is
to check the phase picked up by a $c$-fermion when it moves around a
plaquette. Turning on a plaquette is done by combination of hopping
operators yielding the phases as $B_f^\bbc$, $B_f^\bc$ and $B_f^c$
that are consistent with \eqref{BBB}. However, this is not generic
for all 2-colexes. For 2-colexe plaquettes that the number of their
edges is a multiple of four, we see that the ground state carries
fluxes. Perhaps the most important of such lattices is the so called
4-8-8 lattice shown in Fig.\ref{foureight}. It contains inner
octagons and squares. Once degenerate perturbation theory is applied
about the strong limit of the system, the effective color code
Hamiltonian in terms of plaquette operators in \eqref{Bs} at 12th
order of perturbation is produced, as follows \bea
\label{effective_488}
H_{eff}&=&-\sum_{f}(k_xB_f^{x}+k_yB_f^{y}+k_zB_f^{z})+\mathrm{multiplaquette~terms},\eea
where sum runs over all squares and octagons.
\begin{figure}
\begin{center}
\includegraphics[width=10cm]{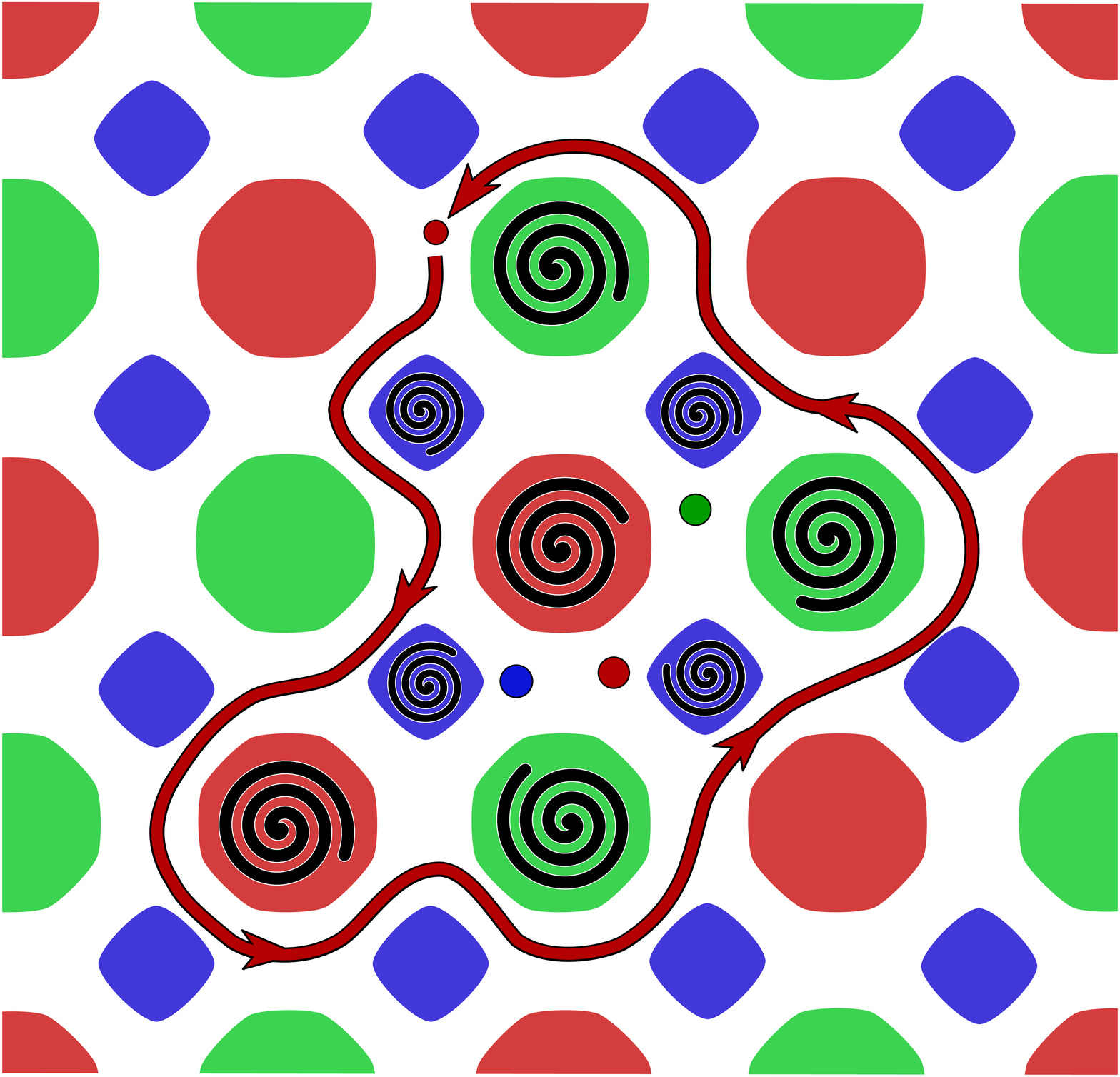} \caption{(color online) A $c$-fermion
process. A red fermion goes around a region $\lambda$. The phase
that it picks up depends on vortex configurations shown by spiral
lines, blue and green fermions and number of plaquettes that the
number of their edges is of multiple four.} \label{gauge}
\end{center}
\end{figure}
Notice that at 12th order of perturbation multiplaquette terms that
are product of square plaquette operators are also appeared. It is
simple to check that the coefficients $k$'s have positive sign. As
we can relate the plaquettes to the representation of gauge group,
the ground state corresponds to vortex free sector. In fact, the
ground state of all 2-colexes with plaquettes of any shape is vortex
free and correspond to $\chi_c=1$ of gauge group. What is able to
differentiate between ground states of 2-colex plaquttes with $4n$
($n$ an integer) edges from others is related to the gauge fields
attached to a fermion. In particular, there is a background of
$\pi$-fluxes in the ground states of such lattices, and the emerging
$c$-fermions can detect them. To make sense of the existence of such
fluxes, let us consider a simple fermionic process as explained
above. When a $c$-fermion turns on a plaquette, the combination of
hopping terms yield $-B_f^\bbc$, $-B_f^\bc$ and $-B_f^c$, which
clearly imply that \bea
\label{flux}(-B_f^\bbc)(-B_f^\bc)(-B_f^c)=-1.\eea This result
exhibit that the ground states of color code models defined on
lattices with $4n$-plaquettes carry fluxes. Thus, in its
representation in terms of $\mathbf{Z}_2\times \mathbf{Z}_2$ gauge
group, the fluxes must be subtracted away.

Our derivation for ground states with $0$- and $\pi$-flux can be
compared with Lieb theorem\cite{lieb_flux}, which states for a
square lattice the energy is minimized by putting $\pi$ flux in each
square face of the lattice. The connection to our models makes sense
when we consider how 2-colexes with hexagonal and 4-8-8 plaquettes
can be constructed from a square lattice by removing some edges.
Then the total $\pi$ fluxes in a set of square faces corresponding
to a 2-colex plaquette amounts to the flux that it carries. It is
simple to see that each hexagonal plaquette composed of two
(imaginary) square faces, and flux $\pi$ in each square face then
implies flux $0$ in the hexagon. The same strategy holds for fluxes
carried by the 4-8-8 plaquette(and in general for all
$4n$-plaquettes). Once again we see that each plaquettes of latter
2-colxes is composed of odd number of square faces, thus they carry
flux $\pi$ in their plaquettes.

Now we can give a general expression for the gauge fields seen by
emerging high-energy fermions. To do so, let us consider a process
in which a $c$-fermion is carried around a region $\lambda$, as in
Fig.\ref{gauge}. The hopping process yields a phase
\begin{equation}\label{phase}
\phi^c_\lambda=\chi_c (q_\lambda) \,
(-1)^{n_\bc^\lambda+n_\bbc^\lambda+n_4^\lambda}
\end{equation}
where $q_\lambda=\prod_\lambda q_f$, $n_c^\lambda$ denotes the
number of $c$-fermions inside $\lambda$ and $n_4^\lambda$ the number
of 2-colex plaquettes inside $\lambda$ with a number of edges that
is a multiple of four. Thus, we can see that each family of fermions
carries a different representation of gauge group given by values of
$q_f$ inside the region. Moreover, it emphasizes that fermions with
different color charges have mutual semionic statistics. Clearly for
hexagonal lattices $n_4^\lambda=0$, and the ground state carries no
fluxes.

\section{Fermionic Mapping}
\label{sect_VI}
\begin{figure}
\begin{center}
\includegraphics[width=9cm]{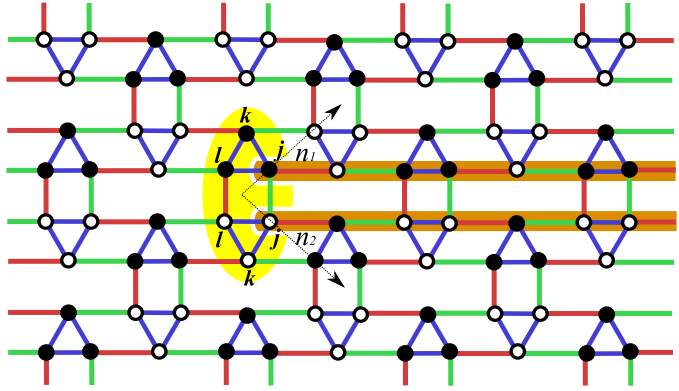} \caption{(color online) Brick-wall lattice that is topologocally
equivalent to the original lattice in Fig.\ref{ruby_lattice}.
Triangles with black and white vertices are chosen such that at the
logical level they correspond to a bipartite lattice. A unit cell of
the lattice has been shown by yellow region and the whole lattice
can be constructed by translation vectors $\mathbf{n_{1}}$ and
$\mathbf{n_{2}}$. The chains supporting the field operator $F$ have
been shown by brown ribbons.} \label{brick_wall}
\end{center}
\end{figure}

In this section we will come back to the original Hamiltonian of
\eqref{2-bodyTCC} on the lattice in order to use another approximate
method based on fermionic mappings. This Hamiltonian can be
fermionized by Jordan-Wigner transformation \cite{Feng_et_al07}. To
do so, firstly it is convenient to present a lattice which is
topologically equivalent to the lattice of Fig.\ref{ruby_lattice}.
This is a new type of ''brick-wall'' lattice as shown in
Fig.\ref{brick_wall}. The black and white sites are chosen such
that, at the effective level, the lattice be a bipartite lattice,
since the effective spins are located at the vertices of hexagonal
lattice representing a bipartite lattice. Note that neither original
lattice in Fig.\ref{ruby_lattice} nor the brick-wall one in
Fig.\ref{brick_wall} are bipartite in their own. Also as we will
see, the fermionization of the model needs some ordering of sites in
the ''brick-wall'' lattice. The unit cell of the brick-wall lattice
is comprised of two triangles as shown in Fig.\ref{brick_wall} in a
yellow ellipse. The translation vectors $\vec{n}_{1}$ and
$\vec{n}_{2}$ connect different unit cells of the lattice.

\begin{figure}
\begin{center}
\includegraphics[width=8cm]{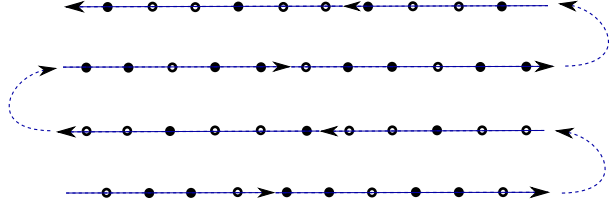} \caption{(color online)
Ordering of sites in the brick-wall lattice of Fig.\ref{brick_wall}.
Such ordering makes it possible to perform a one-dimensional
Jordan-Wigner transformation.} \label{ordering_brick}
\end{center}
\end{figure}

The deformation of the original lattice into a ''brick-wall''
lattice allows one to perform the one dimensional Jordan-Wigner
transformation. The one dimensional structure of the lattice is
considered as an array of sites on a contour as shown in
Fig.\ref{ordering_brick}. The sites on the contour can be labeled by
a single index and the ordering of the sites is identified by the
direction of the arrows in Fig.\ref{ordering_brick}. The expression
of the Pauli operators in terms of spinless fermions will be: \bea
\label{jordan_wigner}
\sigma^{+}_{j}=a^{\dag}_{j}\exp\left(i\pi\sum_{l<j}a^{\dag}_{l}a_{l}\right),
~~~ \sigma^{z}_{j}=(2a^{\dag}_{j}a_{j}-1), \eea where spinless
fermions satisfy the usual anticommutation relations as follows \be
\{a_{i},a^{\dagger}_{j}\}=\delta_{ij},~~\{a^{\dagger}_{i},a^{\dagger}_{j}\}=\{a_{i},a_{j}\}=0.\ee
Next, we introduce Majorana fermions as follows: \bea
\label{majorana1}
c_{j}=-i(a^{\dag}_{j}-a_{j}),~~~d_{j}=a^{\dag}_{j}+a_{j} \eea for
black sites and \bea \label{majorana2}
c_{j}=a^{\dag}_{j}+a_{j},~~~d_{j}=-i(a^{\dag}_{j}-a_{j}) \eea for
white sites. Majorana operators are Hermitian and satisfy the
following relations: \be
\label{majorana_commutation}k^{2}_{j}=1,~~~k_{j}k_{i}=-k_{i}k_{j}~~~i\neq
j,\ee where $k=c,d$.
For each unit cell we can realize two vertical red and green links
corresponding to $\sigma^{x}\sigma^{x}$ and $\sigma^{y}\sigma^{y}$
interactions. Observe that each vertex of the lattice can be
specified by three indices as follows: (i) index $\mathrm{I}$ is
introduced to specify which unite cell the vertex belongs to, (ii)
$\mathrm{b}$ ($\mathrm{w}$) denotes the black (white) vertex, (iii)
$l$, $j$ or $k$ label the position of the vertex in each triangle.
With this labeling of vertices, the expression of the different
terms appearing in the two-body Hamiltonian in terms of Majorana
fermions is listed below: \bea \label{mapping_fermionization} &&
\mathrm{blue~
links}:~\sigma^{z}_{j}\sigma^{z}_{k}=i\hat{U}_{jk}d_{j}d_{k},\nonumber\\
&& \mathrm{horizontal~r-links}:~\sigma^{x}_{j}\sigma^{x}_{k}=id_{j}d_{k},\nonumber \\
&&\mathrm{horizontal~g-links}:~\sigma^{y}_{j}\sigma^{y}_{k}=-id_{j}d_{k},\nonumber\\
&&\mathrm{vertical~r-links}:~\sigma^{(\mathrm{Iw})x}_{l}\sigma^{(\mathrm{Ib})x}_{l}=id^{(\mathrm{Iw})}_{l}d^{(\mathrm{Ib})}_{l}\hat{F}_{\mathrm{I}},\nonumber\\
&&\mathrm{vertical~g-links}:~\sigma^{(\mathrm{Iw})y}_{j}\sigma^{(\mathrm{Ib})y}_{j}=-id^{(\mathrm{Iw})}_{j}d^{(\mathrm{Ib})}_{j}\hat{F}_{\mathrm{I}},\eea
where $\hat{U}_{jk}=-ic_{j}c_{k}$ and the operator
$\hat{F}_{\mathrm{I}}$ is a non-local operator. Interestingly
enough, this non-local term has the following expression: \bea
\label{F1}\hat{F}_{\mathrm{I}}=\exp(i\pi\sum^{k-1}_{l=j}a^{\dag}_{l}a_{l})=\prod_{l\in
R}ic_{l}d_{l}, \eea where we have used the ordering of the
brick-wall lattice and $R$ stands for a set of spins crossed by two
\emph{brown} ribbons shown in Fig.\ref{brick_wall}.
Observe that for each unit cell we can realize such ribbons. The
non-local operator can be written as a product of some plaquette
operators each corresponding to a rectangle(inner hexagon) wrapped
by the ribbons. Notice that an inner hexagon of the ruby lattice in
Fig.\ref{ruby_lattice} looks like a rectangle in the brick-wall
lattice. Thus, two ribbons are nothing but a combination of
rectangles as in Fig.\ref{brick_wall}. This plaquette operator
precisely corresponds to operator $S^{C}_{f}$ in
Fig.\ref{plaquette_IOM_Independent}. Let $f_{1}, f_{2},...$ stand
for those rectangles (inner hexagons). Thus, we have: \bea
\label{F2} \hat{F}_{\mathrm{I}}=\prod_{l\in
R}c_{l}d_{l}=S^{C}_{f_{1}}\otimes S^{C}_{f_{2}}\otimes
S^{C}_{f_{3}}....\eea After all these transformations, we arrive at
an exact fermionized Hamiltonian given by \bea
\label{fermionized_Hamiltonian}
H&=&-J_{z}\sum_{b-link}i\hat{U}_{jk}d_{j}d_{k}
-J_{x}\sum_{r-link}id_{j}d_{k}+J_{y}\sum_{g-link}id_{j}d_{k}\nonumber
\\&&-J_{x}\sum_{\mathrm{I}}id^{(\mathrm{Iw})}_{l}d^{(\mathrm{Ib})}_{l}\hat{F_{\mathrm{I}}}+J_{y}\sum_{\mathrm{I}}id^{(\mathrm{Iw})}_{j}d^{(\mathrm{Ib})}_{j}\hat{F}_{\mathrm{I}}.\eea
It is simple to check the following commutation relations:
\bea\label{commuting_fildes}
\left[H,\hat{F}_{_{I}}\right]=\left[H,\hat{U}_{jk}\right]=\left[\hat{F}_{\mathrm{I}},\hat{F}_{\mathrm{I}'}\right]=0.\eea

Since $\hat{F}_{\mathrm{I}}^{2}=1$, the nonlocal operators can be
replaced by its eigenvalues $F_{\mathrm{I}}=\pm 1$ . Thus the
nonlocal terms appearing in the Hamiltonian which are related to the
vertical links of the brick-wall lattice can be reduced to quadratic
terms. However, the first sum in \eqref{fermionized_Hamiltonian}
which is related to the triangles cannot be reduced to the quadratic
term. This is because local fields $\hat{U}_{jk}$ corresponding to
three links of a triangle anticommute with each other as well as
with some $\hat{F}_{\mathrm{I}}$'s. Due to these anticommutation,
all fields cannot be fixed independently. This fact is in sharp
contrast with Kitaev's model and its variants. These latter models
are defined on trivalent lattices. The different fields live at
spatially separated links allowing a free fermion exact solution
\cite{honeycomb,Feng_et_al07}. The obtained quadratic Hamiltonian
describes free majorana fermions in a background of $\mathbf{Z}_{2}$
charges. Instead, the lattice of our model is 4-valent, a sharp
difference that prevents complete solvability and gives rise to very
interesting features not present in the mentioned models. Note that
the fields $\hat{U}_{jk}$ are highly interacting since on a triangle
three fields go to vacuum in the sense that
$\hat{U}_{jk}\hat{U}_{kl}\hat{U}_{lj}=-1$. This is resemblance of
vertex interaction in high energy fermions that we have seen in
Sect.\ref{sect_V} with the bosonic mapping. However, this latter
relation doesn't coincide with the symmetry of the model as they do
not commute with each other. This will be considered next.

Thus far, we have considered fields that are present in the
Hamiltonian. In what follows, we introduce another set of fields
which have the $\mathbf{Z}_{2}\times\mathbf{Z}_{2}$ symmetry
commuting with each other and with the Hamiltonian. To this end,
consider a plaquette $f$. As before, by a plaquette we mean an outer
and an inner hexagon with six triangles between them. Let $V_{f}$
and $V_{h}$ stand for sets of vertices of plaquette and inner
hexagon, respectively. It is natural that $V_{h}\subset V_{f}$. To
each plaquette we attach the three following fields: \bea
\label{fermion_fildes} \phi^{1}_{f}=\prod_{j\in
V_{f}}c_{j},~\phi^{2}_{f}=\prod_{j\in f\backslash h}c_{j}
\prod_{v\in V_{h}}d_{j},~\phi^{3}_{f}=\prod_{j\in
V_{h}}c_{j}d_{j},\eea where by $f\backslash h$ we simply mean
$V_{f}-V_{h}$. Each $\phi_{f}$ squares identity. They commute with
each other and with Hamiltonian and $\hat{F_{\mathrm{I}}}$: \bea
\left[\phi^{k}_{f},\phi^{k'}_{f'}\right]=\left[\phi^{k}_{f},\hat{F}_{\mathrm{I}}\right]=\left[H,\phi^{k}_{f}\right]=0
\eea Also, the fields $\phi_{f}$ are responsible of the
$\mathbf{Z}_{2}\times\mathbf{Z}_{2}$ gauge symmetry since
$\phi^{1}_{f}\otimes \phi^{2}_{f} \otimes\phi^{3}_{f}=-1$. The above
fields are related to the plaquette operators.Using the
transformations we have introduced in
\eqref{mapping_fermionization}, we can fermionize the conserved
plaquette operators obtained in Fig.\ref{plaquette_IOM_Independent}.
They are associated to the above constants of motion as follows \bea
\label{plaquette_fermionization}
S^{A}_{f}=\phi^{1}_{f}\hat{F}_{\mathrm{I}},~~~
S^{B}_{f}=\phi^{2}_{f}\hat{F}_{\mathrm{I}},~~~
S^{C}_{f}=\phi^{3}_{f}.\eea Although the above gauge fields make it
possible to divide the Hilbert space into sectors in which be
eigenspaces of gauge fields (or eigenspaces of plaquette operators),
they do not allow us to reduce the Hamiltonian in
\eqref{fermionized_Hamiltonian} into a quadratic form. The
$\hat{F}_{\mathrm{I}}$'s can be fixed as they commute with the
Hamiltonian. But, we are not able to reduce the four-body
interaction terms in the Hamiltonian into quadratic form. In fact,
the anticommutation of $\hat{U}_{jk}$'s on a blue triangle prevents
them to be fixed consistently with gauge fixing.
\section{Conclusions}
\label{sect_VII}
We have introduced a two-body spin-1/2 model in a ruby lattice, see
Fig.\ref{ruby_lattice}. The model exhibits an exact topological
degeneracy in all coupling regimes. The connection to the
topological color codes can be discussed on the non-perturbative
level as well as confirmed by perturbative methods. In the former
case, on the ruby lattice we realized plaquette operators with local
$\mathbf{Z}_{2}\times\mathbf{Z}_{2}$ symmetry of the color codes.
All plaquette operators commute with the Hamiltonian and they
correspond to integrals of motion. The plaquettes can be extended to
more complex objects that can be considered as string-nets:
non-trivial strings with branching points. The nontrivial strings
corresponding to the various homology classes of the manifold
determine the exact degeneracy of the model. For the case of
periodic boundary conditions, i.e on a torus, and for each
non-contractible cycle of the torus, we can identify three
nontrivial closed strings. Once each of them is colored, the
plaquettes of the lattice can be correspondingly colored as in
Fig.\ref{lattice_letter}. For each homology class, they are related
to each other by the gauge symmetry of the model. The crucial
property of these strings is that they commute with Hamiltonian but
not always with each other. This is independent of the regimes of
coupling constants of the model. Being anticommuting closed
nontrivial strings, the model has exact topological degeneracy. To
clarify this observation, we use perturbation theory to investigate
a regime of coupling corresponding to a strong coupling limit
(triangular limit). In this limit the topological color code will be
the effective description of the model. The effective representation
of the closed loop operators determine the terms appearing in the
effective Hamiltonian at different orders.

Unlike the Kitaev's model or any its variants, our model is not
integrable in terms of mapping to Majorana fermions, to the best of
our knowledge. This model is a four-valent lattice and gauge fields
not always commute with each other. However, we have emphasized in
Sect.\ref{sect_III} that the existence of exact integrals of motion
(IOMs) at a non-perturbative level is far more enriching than
demanding exact-solvability of a model. In fact, if the number of
IOMs is large enough, the model can turn out to be solvable. Thus,
fixing plaquette operators can not give rise to fix all gauge
fields.

The description of our model in terms of hard core bosons yields
very fruitful and interesting physics of the model. Using a bosonic
mapping, it is possible to discuss the emergence of strongly
interacting anyonic fermions. They form three classes each of one
color. Fermions from different classes have mutual semionic
statistics. A very intriguing feature of these fermions is related
to the topological color charges they carry. They carry charges from
a particular family of low energy fermions. Thus the charges created
by open strings are invisible to high energy fermions. Moreover,
there are some experimental proposals to realize hard-core bosons
with optical lattices\cite{experiment_hardcore} and it would be a
nice challenge to implement a Hamiltonian like \eqref{Hamiltonian_B}
and \eqref{Terms}.

We have shown that this new model exhibits enough novel interesting
and relevant properties so as to justify further research. Some of
these possible lines of study are as follows: We have only studied a
particular phase of the system, although we are able to study
non-perturbative effects as well. The fact that all phases show a
topological degeneracy anticipates a rich phase diagram. In this
regard, one may explicitly break the color symmetry that the model
exhibits and still keep the features that we have discussed. It
would be particularly interesting to check whether any of the phases
displays non-abelian anyons. The model has many integrals of motion,
although not enough to make it exactly solvable. This becomes
another appealing feature of the model since other methods of study,
like numerical simulations and experimental realizations will help
to give a complete understanding of all its
phases.\\\\

\noindent {\em Acknowledgements} We acknowledge financial support
from a PFI grant of EJ-GV, DGS grants under contracts,
FIS2006-04885, and the ESF INSTANS 2005-10.

\appendix

\section{2-Body Hamiltonian for Color Codes using Cluster States}
\label{sect_appendix}
\begin{figure}
\begin{center}
\includegraphics[width=12cm]{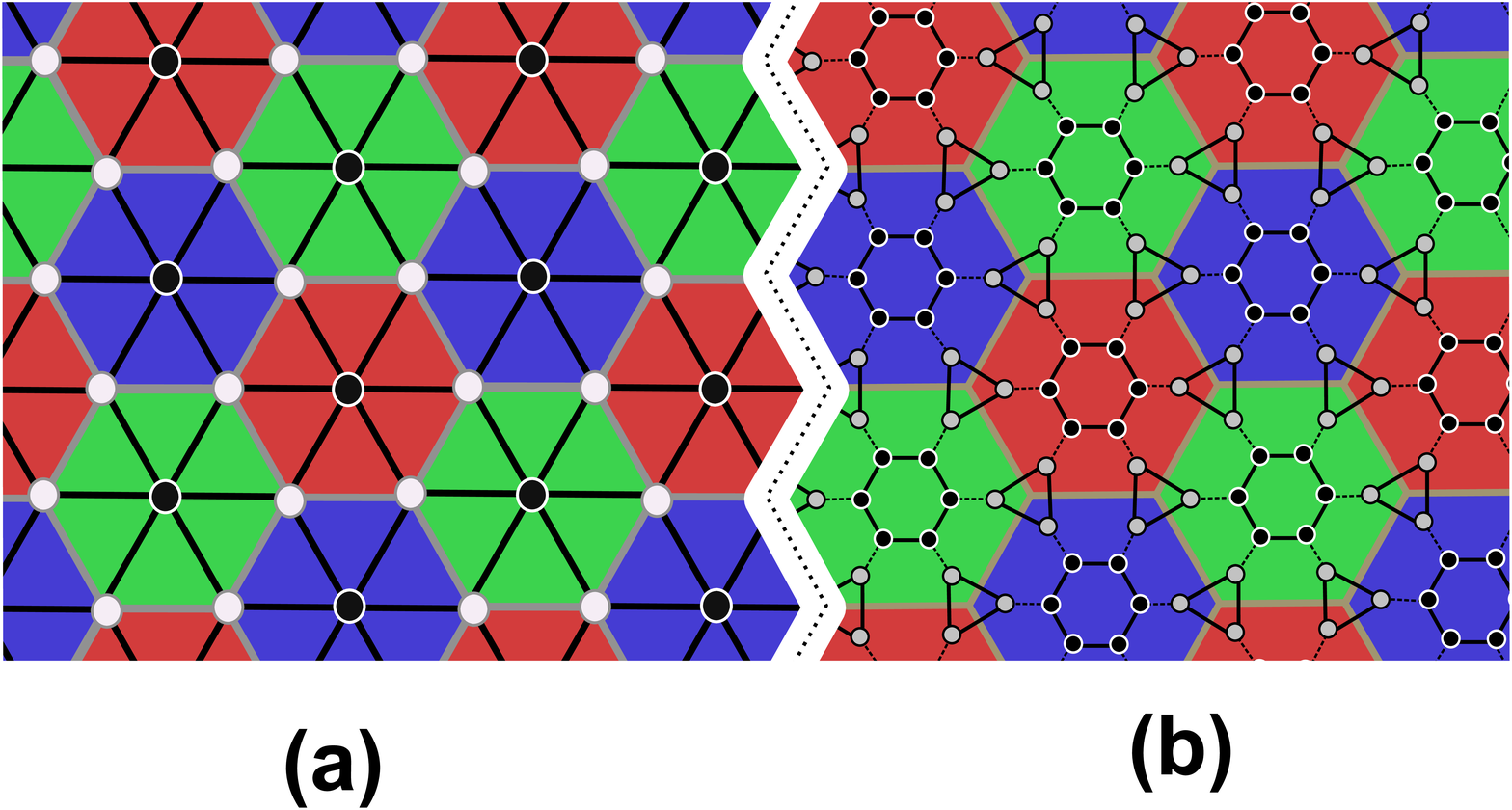} \caption{(color online) (a)
The graph needed for obtaining color codes from graph states. The
graph is bipartite. Black and white vertices correspond to the
plaquettes and vertices of a 2-colex. Black solid links are edges of
the graph (b) The corresponding graph state can be approximated as a
low energy description of a lattice with 2-body Hamiltonian. The
lattice is obtained from the graph by replacing its vertices with
some hexagons and triangles. The interactions $\sigma^{z}\sigma^{z}$
and $\sigma^{x}\sigma^{z}$ are associated to the solid and dashed
links, respectively.} \label{graph}
\end{center}
\end{figure}

A topological color code can be constructed from a  graph state
defined on a bipartite lattice by means of a set of measurements on
certain subsystems. This bipartite lattice is shown in
Fig.\ref{graph}(a), where the black vertices correspond to
plaquettes and the white vertices correspond to the vertices of a
2-colex. To this graph we can attach a set of stabilizers as follows
\bea \label{stabilizer_graph} K_{\alpha}=X_{\alpha}\prod_{\prec
\alpha, \beta\succ}Z_{\beta},\eea where $\alpha$ and $\beta$ stand
for vertices of the graph and the product runs over all vertices
that are connected to $\alpha$ by black links. Let us set
$V=\mathfrak{U}_{1}\cup \mathfrak{U}_{2}$, where $\mathfrak{U}_{1}$
and $\mathfrak{U}_{2}$ stand for the set of white and black vertices
of the bipartite graph in Fig.\ref{graph}(a). Note that white and
black vertices corresponds to the vertices and plaquettes of the
2-colex. This bipartite graph is exactly what we need to construct
color codes. To this end, we first impose a unitary transformation
on the sublattices that allows us to have a more symmetric form of
the stabilizer operators, i.e
\bea \no \forall~~ v\in \mathfrak{U}_{1}~~~~~K_{v}=X_{N(v)} \\
\forall~~ f\in \mathfrak{U}_{2}~~~~~K_{f}=Z_{N(f)},\eea where $N(v)$
denotes the site $v$ and its neighbors, and the same goes for
$N(f)$. The corresponding cluster state denoted by $\ket{G}$ will be
the common eigenvector of the above stabilizer operators. Thus we
have: \bea \no \forall~~ v\in
\mathfrak{U}_{1}~~~~~X_{N(v)}|G\rangle=|G\rangle \\
\label{clusterstate} \forall~~ f\in
\mathfrak{U}_{2}~~~~~Z_{N(f)}|G\rangle=|G\rangle. \eea Finally, a
graph state can be related to a color code within a set of
measurements in the $Z$ basis on all qubits corresponding to the set
$\mathfrak{U}_{2}$.

We suppose there is a two dimensional lattice of physical qubits
that is governed by a 2-body Hamiltonian. Physical qubits of the
lattice are projected to \emph{logical} qubits. The point is that
this projection is achieved by going to some order in perturbation
theory. We think of vertices of the graph in Fig.\ref{graph}(a) as
logical qubits. The lattice with 2-body interaction is shown in
Fig.\ref{graph}(b), where the number of physical qubits
corresponding to the vertices of the graph equals the number of
links crossing the vertex. The new resulting lattice consists of
\emph{triangles} and \emph{hexagons} and physical qubits live on
their vertices. Triangles and hexagons are in one to one
correspondence with the white and black vertices ($\mathfrak{U}_{1}$
and $\mathfrak{U}_{2}$) of the graph in Fig.\ref{graph}(b),
respectively. Note that each triangle is linked with three
neighboring hexagons and each hexagon is linked with six neighboring
triangles.

The low lying spectrum of a well constructed 2-body Hamiltonian
defined on the lattice composed of hexagons and triangles may
describe a cluster state. To this end, we need for the following
projection from the physical qubits to the logical ones: \bea
\label{projection_graph}
P_{v}=|\!\Uparrow_{v}\rangle\langle\uparrow\uparrow\uparrow\!|+|\!\Downarrow_{v}\rangle\langle\downarrow\downarrow\downarrow\!|,
~~P_{f}=|\!\Uparrow_{f}\rangle\langle\uparrow\uparrow\uparrow\uparrow\uparrow\uparrow\!|+|\!\Downarrow_{f}\rangle\langle\downarrow\downarrow\downarrow\downarrow\downarrow\downarrow\!|
\eea where $|\!\Uparrow_{L}\rangle$ and $|\!\Downarrow_{L}\rangle$
with $L=v,f$ stand for the two states of the logical qubits obtained
within the above projections, or alternatively they are states of
logical qubits of the graph in Fig.\ref{graph}(a). We set the
following Hamiltonian: \bea \label{Hamiltonian_graph}H=H_{0}+\lambda
V \eea where $H_{0}$ is the unperturbed Hamiltonian which can be
treated exactly and $\lambda$ is a small quantity which allow us to
treat the term $\lambda V$ perturbatively. We refer to each triangle
(hexagon) and its vertices by a site index $v$ ($f$) and indices
$i,j$, respectively. The unperturbed part of the Hamiltonian
included in $H_{0}$ is as follows. \bea \label{unperurbed_graph}
H_{0}=-\sum_{L}\sum_{<i,j>}\sigma^{z}_{L,i}\sigma^{z}_{L,j} \eea
where the first sum runs over all triangles and hexagons (sites) and
$<\!\!i,j\!\!>$ stand for the nearest-neighbor qubits around the
corresponding triangle or hexagon connected by the solid lines as in
the Fig.\ref{graph}(b). The interaction between qubits of triangles
and qubits of hexagons are included in $V$: \bea
\label{interaction_graph} V=-\sum_{<v i, f
j>}(\sigma^{x}_{v,i}\sigma^{z}_{f,j}+\sigma^{z}_{v,i}\sigma^{x}_{f,j})
\eea where $<\!\!v i, f j\!\!>$ denotes two neighboring qubits which
are connected by the dashed line. The unperturbed Hamiltonian
$H_{0}$ can easily be  diagonalized for triangles and hexagons.
Ground state vectors of a triangle or hexagon form a two dimensional
space spanned by the following polarized eigenvectors: \bea
\label{polirizedtri_graph} \mathrm{triangle:}
~~|\!\Uparrow_{v}\rangle\equiv|\!\uparrow\uparrow\uparrow\rangle,~~~|\!\Downarrow_{v}\rangle\equiv|\!\downarrow\downarrow\downarrow\rangle
\eea and \bea \label{polirizedhexa_graph} \mathrm{hexagon:}
~~|\!\Uparrow_{f}\rangle\equiv|\!\uparrow\uparrow\uparrow\uparrow\uparrow\uparrow\rangle,~~~|\!\Downarrow_{f}\rangle\equiv|\!\downarrow\downarrow\downarrow\downarrow\downarrow\downarrow\rangle.
\eea Let $N_{v}$ and $N_{f}$ be the number of triangles and
hexagons, respectively. Imposing periodic boundary conditions, the
total number of triangles and hexagons will be:
$N=N_{v}+N_{f}=3N_{f}$ . Thus, the dimension of the ground-space of
the unperturbed Hamiltonian $H_{0}$ (or ground state degeneracy)
becomes: $g_{0}=2^{N_{v}}2^{N_{f}}=2^{N}$, and the ground state
energy is: $E_{0}^{(0)}=N_{f}(-6)+N_{v}(-3)=-4N$ in terms of the
energy scale of the problem. The first excited state is produced by
exciting one of the triangles or hexagons and has energy:
$E_{1}^{(0)}=N_{f}(-6)+N_{v}(-3)+4=-4(N-1)$ with degeneracy
$g_{1}=14N2^{N}$. The second excited state has energy
$E_{1}^{(0)}=N_{f}(-6)+N_{v}(-3)+8=-4(N-2)$ with degeneracy:
$g_{2}=4(N+5N^{2})2^{N-1}$, and so on and so forth.

Using degenerate perturbation theory as in Sec.\ref{Green_Function},
the effect of perturbation $V$ on the ground sate subspace can be
investigated, and see if it breaks the degeneracy. It is simple to
see that first order perturbation does not have anything to do with
the ground state subspace. The second order gives rise to a trivial
effect as a shift in energy, since each operator related to dashed
links appears twice. The third order perturbation theory however
gives rise to a nontrivial effect. It causes a partial lift of the
ground state degeneracy, but not complete. The initial degeneracy
$2^{N}$ gets reduced down to  $2^{N_{f}}$. This nontrivial effect
arises from the product of three dashed links crossing a typical
triangle, namely the ground state vectors are grouped into the
$2^{N_{v}}$ states, each containing $2^{N_{f}}$ vectors. The product
of three (six) $\sigma^{x}$ operators around a triangle (hexagon) is
equivalent to an $X$ operator acting on the logical qubit which is
projected down from the three(six) qubits of the triangle(hexagon) ,
since \bea \label{xlogical_graph}
X_{L}=|\!\Uparrow_{L}\rangle\langle\Downarrow_{L}\!|+|\!\Downarrow_{L}\rangle\langle\Uparrow_{L}\!|.\eea
Also the action of a $\sigma^{z}$ on one qubit of a triangle or
hexagon is equivalent to an $Z$ operator acting on the related
logical qubit, since \bea \label{zlogical_graph}
Z_{L}=P_{L}\sigma^{z}P^{\dagger}_{L}=|\!\Uparrow_{L}\rangle\langle\Uparrow_{L}\!|-|\!\Downarrow_{L}\rangle\langle\Downarrow_{L}\!|.\eea
Now we can go on in order to calculate the third order perturbation:
\bea \label{effective3_graph}
H^{(3)}_{\mathrm{eff}}=-\frac{3!}{(E^{(0)}_{0}-E^{(0)}_{1})^{2}}\sum_{v}K_{v}=-\delta\sum_{v}K_{v},
\eea where \bea \label{stabilizertri_graph}
K_{v}=X_{v}\prod_{f}Z_{f}, \eea the product runs over three black
vertices linked to the $v$ and $\delta=\frac{3}{8}$. The operator
$K_{v}$ is a stabilizer for the logical qubits which are projected
down from the triangles. Since $K_{v}^{2}=1$, the ground states
correspond to the values of $k_{v}=+1$. We skip the forth and fifth
order of perturbation because they have trivial effects.

Like in $3^{rd}$ order perturbation, we are faced with a nontrivial
term in the $6^{th}$ order perturbation theory. We will see that by
considering this order, the ground state degeneracy is lifted
completely. This nontrivial effect arises from the product of terms
in the perturbation $\lambda V$ corresponding to the links around a
hexagon. Finally, for the $6^{th}$ order perturbation we have: \bea
\label{effective6_graph}
H^{(6)}_{\mathrm{eff}}=-\gamma\sum_{f}K_{f},\eea where\bea
\label{stabilizerhexa_graph}K_{f}=X_{f}\prod_{v}Z_{v},\eea and the
product runs over six white vertices linked to the hexagon $f$. The
coefficient $\gamma$ has positive sign and its precise value is
unimportant. We would like to emphasize that at six order in
perturbation theory some other terms appear which are product of two
distinct $K_{v}$. However, we skip them as they all commute.
Equations \eqref{stabilizertri_graph} and
\eqref{stabilizerhexa_graph} provides all we need to adopt the
cluster state in \eqref{clusterstate} as ground state of the low
energy effective theory of Hamiltonian in \eqref{Hamiltonian_graph},
which up to six order of perturbation can be written as follows \bea
 H_{\mathrm{eff}}=\mathrm{constant}
-\delta\sum_{v}K_{v}-\gamma\sum_{f}K_{f}.\eea We see that the above
effective Hamiltonian is completely different from that of in
\eqref{effective_Hamiltonian}. The latter equation gives rise
directly to the topological color code as its ground state, but the
ground state (cluster state) of former one needs further local
measurements to encode the desired color code.


\end{document}